\documentclass[preprint,journal]{vgtc}                % final 
\ifpdf%                                % if we use pdflatex
  \pdfoutput=1\relax                   % create PDFs from pdfLaTeX
  \usepackage{graphicx}                % allow us to embed graphics files
  \DeclareGraphicsExtensions{.pdf,.png,.jpg,.jpeg} % for pdflatex we expect .pdf, .png, or .jpg files
\else%                                 % else we use pure latex
  \usepackage{graphicx}                % allow us to embed graphics files
  \DeclareGraphicsExtensions{.eps}     % for pure latex we expect eps files
\fi%

\graphicspath{{figures/}{pictures/}{images/}{./}} % where to search for the images

\usepackage{microtype}                 % use micro-typography (slightly more compact, better to read)
\PassOptionsToPackage{warn}{textcomp}  % to address font issues with \textrightarrow
\usepackage{textcomp}                  % use better special symbols
\usepackage{mathptmx}                  % use matching math font
\usepackage{times}                     % we use Times as the main font
\usepackage{cite}                      % needed to automatically sort the references
\usepackage{tabu}                      % only used for the table example
\usepackage{booktabs}                  % only used for the table example
\usepackage{amsmath}
\usepackage{amssymb}
\usepackage{tabulary}
\usepackage{multirow}
\usepackage{enumerate}
\onlineid{0}

\vgtccategory{Research}
\vgtcpapertype{application/design study}

\newcommand{\toolname}{AttnAnalyzer\xspace}

\usepackage{tikz}

\title{A Visual Analytics System for Improving Attention-based Traffic Forecasting Models}

\author{
Seungmin Jin$^*$, Hyunwook Lee$^*$, Cheonbok Park, Hyeshin Chu, Yunwon Tae, Jaegul Choo, and
Sungahn Ko$^{**}$}
\authorfooter{
\item[]{$^{*}$Equal contribution}
\item[]{$^{**}$Corresponding author}
\item Seungmin Jin, Hyunwook Lee, Hyeshin Chu, and Sungahn Ko are with UNIST. E-mail: \{skyjin, gusdnr0916, hyeshinchu, sako\}@unist.ac.kr
\item Cheonbok Park is with NAVER. E-mail: cbok.park@navercorp.com
\item Yunwon tae and Jaegul Choo are with KAIST. E-mail: \{tyj204, jchoo\}@kaist.ac.kr
}

\shortauthortitle{Jin and Lee \MakeLowercase{\textit{et al.}}: A Visual Analytics System for Improving Traffic Forecasting Models}

\abstract
{With deep learning (DL) outperforming conventional methods for different tasks, much effort has been devoted to utilizing DL in various domains. 
Researchers and developers in the traffic domain have also designed and improved DL models for forecasting tasks such as estimation of traffic speed and time of arrival. 
However, there exist many challenges in analyzing DL models due to the black-box property of DL models and complexity of traffic data (i.e., spatio-temporal dependencies). 
Collaborating with domain experts, we design a visual analytics system, \toolname, that enables users to explore how DL models make predictions by allowing effective spatio-temporal dependency analysis. 
The system incorporates dynamic time warping (DTW) and Granger causality tests for computational spatio-temporal dependency analysis while providing map, table, line chart, and pixel views to assist user to perform dependency and model behavior analysis. 
For the evaluation, we present three case studies showing how \toolname can effectively explore model behaviors and improve model performance in two different road networks. We also provide domain expert feedback.
}% end of abstract


\teaser{
 \includegraphics[width=0.85\linewidth]{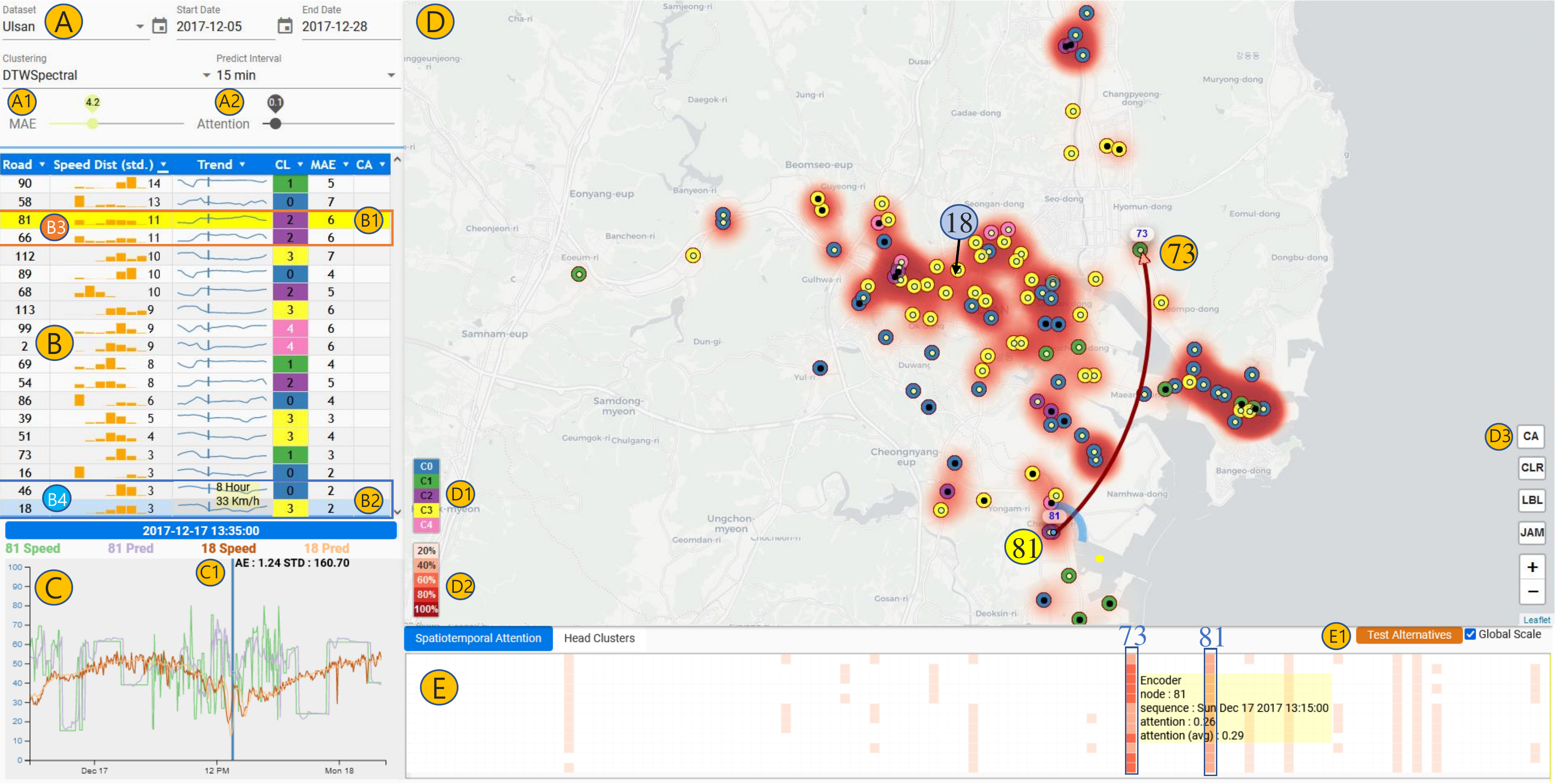}
 \vspace{-0.3cm}
 \centering
  \caption{Our system for analyzing how an attention-based deep learning model predicts traffic congestion. (a) filter view, (b) table view, (c) ground-truth\&prediction result comparison view, (d) a map with attention curves and clusters, and (e) attention heatmap view.}
\label{fig_teaser}
}

\renewcommand{\manuscriptnotetxt}{
Manuscript received 21 Mar. 2022; revised 6 June 2022; accepted 8 Aug. 2022. Date of Publication xx xxx. 202x; date of current version xx xxx. 202x. Digital Object Identifier: xx.xxxx/TVCG.201x.xxxxxxx
}
\vgtcinsertpkg
\usepackage{xspace}
\usepackage{kotex}
\usepackage{microtype}                 % use micro-typography (slightly more compact, better to read)
\PassOptionsToPackage{warn}{textcomp}  % to address font issues with \textrightarrow
\usepackage{textcomp}                  % use better special symbols
\usepackage{mathptmx}                  % use matching math font
\usepackage{times}                     % we use Times as the main font
\usepackage{cite}                      % needed to automatically sort the references
\usepackage{tabu}                      % only used for the table example
\usepackage{booktabs}                  % only used for the table example

\usepackage{tabulary}
\usepackage{multirow}
\usepackage{enumerate}
\usepackage{tikz}
\usepackage{appendix}

\newcommand*{\iconAttn}{\includegraphics[width=0.03\linewidth]{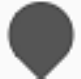}}%
\newcommand*{\iconMAE}{\includegraphics[width=0.03\linewidth]{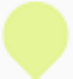}}%

\newcommand{\ars}{AttnArrows\xspace}

\newcommand{\rev}[1]{#1}

\newcommand{\stgrat}{ST-GRAT\xspace}
\newcommand{\ar}{AttnArrows\xspace}
\newcommand{\comment}[1]{}
\newcommand*{\iconTriangle}{\includegraphics[width=0.03\linewidth]{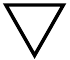}}%

\begin{document}


\maketitle

\section{Introduction}
To mitigate traffic congestion, which is a major issue in modern cities, a lot of effort has been devoted to developing methods for predicting future congestion to support drivers' route decisions~\cite{akhtar2021review}. 
In particular, as many recent studies show that deep learning (DL) outperforms conventional methods in forecasting traffic congestion \cite{tedjopurnomo2020survey} and the estimated time of arrival (ETA)~\cite{yin2021deep}, there have been numerous practical approaches for developing and deploying new DL models. 
For example, reporters in broadcasting centers announce which roads will be congested in the next 15, 30, and 60 minutes to help drivers avoid the possibly congested roads or change their driving schedule (i.e., delayed departure time)~\cite{Lee19}.
Mobility industries, such as navigation service providers, also report that they have achieved greater than 40\% accuracy in estimating travel time of drivers with DL~\cite{derrow2021eta}.

But, many obstacles exist in improving DL models in the traffic domain.
First, DL methods are black-box in nature~\cite{Guidotti18}, which means they do not show what they have learned during training or how they make predictions given input features.  
Second, traffic domain data is spatio-temporal, which is hard to analyze, as both space and time should be considered simultaneously.
For example, when a road is congested because of an accident, there is a high chance that the roads linked to the road with the accident will soon become congested, but it is hard to estimate when the congestion happens and which neighboring roads would be affected~\cite{li2018brief}.
Third, traffic data is heterogeneous with extreme cases~\cite{li2018brief, tedjopurnomo2020survey, yuan2018hetero} and easily affected by uncontrollable external factors (e.g., accidents).
Thus the traffic prediction task is especially challenging in that the models need to learn not only spatio-temporal features from the data, but also how to respond to implicit external events on roads. The external factors could even vary by region~\cite{Lee19}, which further complicates to the models.
Indeed, spatio-temporal analysis with a large space (e.g., major city's road network) and different time units (e.g., hours, days, months) is a nontrivial task~\cite{andrienko2003exploratory}.

In this work, we collaborate with domain experts to design a visual analytics (VA) system that supports users in effectively exploring how a DL model predicts future traffic conditions and in finding insights for performance improvements. 
We first perform task analysis with three domain experts and extract the system requirements needed to support the experts in answering questions that should be answered for their work.
Then, we design a VA system, \toolname, that provides three automated methods and visual interfaces for exploration. 
We incorporate the dynamic time warping (DTW)~\cite{salvador2007toward} and Granger causality tests to support users in exploring spatio-temporal dependencies of roads, such as proceeding and lagging speed patterns. 
We also utilize a spectral clustering algorithm for grouping roads with similar speed patterns for automated methods. 
There are several visualizations in \toolname, including table, line chart, map and pixel views. 
The table view presents information of roads, such as speed trends, cluster index, and prediction accuracy. 
The map view allows spatial dependency analysis, while the line chart view reveals the temporal dependencies of roads. 
The pixel views have two sub-views to analyze attention information, extracted from an attention-based traffic forecasting model, called \stgrat. 
To validate hypotheses created from the system, we devise an attention enforcement method, which replaces the attention of problematic roads with those from the roads with low MAE error.
For demonstration purposes, we use an attention-based traffic forecasting model because they show state-of-the-art performance and because experts have expressed that they extensively use attention-based DL models in navigation service.

To evaluate the system, we provide three case studies. 
In the first two case studies, we showcase how domain experts use the system to explore the spatio-temporal dependencies of two different large road networks. 
In the last case study, we show that the insights derived from the two case studies are meaningful for improving model performance, confirming with attention enforcement. 
Lastly, we provide domain expert feedback and discuss limitations and future work.

The main contributions of this work include the following: 1) a VA system design for exploring traffic forecasting model's behaviors from a spatio-temporal perspective, 2) incorporation of automated methods (DTW, Granger causality test, clustering) for visual temporal analysis, 3) development of an attention enforcement method, and 4) quantitative and qualitative evaluations of the system with three case studies, proven model's accuracy improvements with the attention enforcement method, and domain experts' feedback. 
Although we show how model designers can improve model performance using our tool, the main objective of this work is not to propose a new model, but to demonstrate how to explain deep learning models with attention.
To our knowledge, this work the first attempt to exploring and attention-based traffic forecasting models' prediction process and improving performance in the traffic domain, demonstrating the power of visual analytics approaches~\cite{Thomas05}.
Traffic data is heterogeneous with extreme cases~\cite{li2018brief, tedjopurnomo2020survey, yuan2018hetero} and affected by uncontrollable external factors, such as accidents.
Thus the traffic prediction task is especially challenging in that the models in the domain need to learn not only spatio-temporal features from the data, but also how to respond to implicit external events on roads.
The external factors could even vary by region~\cite{Lee19}, which further complicates to the models.

\section{Related Work}

\subsection{Spatio-Temporal Models for Traffic Prediction}
As deep learning models are effective in many domains with large data sets, researchers have developed various models for the traffic domain. 
In particular, many models have been proposed for speed prediction that can be used in the field to relieve traffic congestion issues~\cite{Lee19}. 
In order to accurately forecast road speeds, most of studies focus on modeling spatial and temporal dependencies that in general affect the dynamics of road speeds~\cite{li2018brief}.
For example, if there is a vehicle accident on a road, it is highly possible that the other roads; linked to that road will become congested (i.e., spatial dependency). 
As time passes, the congestion beginning at the road with the accident may propagate to other roads (i.e., temporal dependency).
As such, existing models aim to effectively model road network dependencies for speed prediction.

To effectively capture spatial dependency, graph convolution neural networks (GCNNs)~\cite{kipf2016semi} has been proposed that apply the convolution technique to graphs of roads.
As using diffusion on graphs turns out to be effective for modeling spatial dependency~\cite{li2018dcrnn} in road networks, many researchers have employed diffusion convolution neural networks (DCNNs) for accurate prediction~\cite{Wu19,li2018dcrnn,Pan2018HyperSTNetHF}.
However, the models based on GCNNs and DCNNs are vulnerable, as they only consider spatial dependency as fixed values, regardless of inputs and traffic condition changes. 
To alleviate this issue, attention-based models have been employed to better model spatial dependency with considerations of road distances, showing state-of-the-art performance for speed prediction tasks~\cite{Zhang2018gaan,park2019stgrat,park2019stgrat}.

Temporal dependency also plays a crucial role in encoding traffic congestion patterns. 
Examples include congestion propagation on roads over time.
To effectively capture the dependency, prior studies have mainly used conventional modeling methods, such as recurrent neural networks (RNNs)~\cite{Zhang2018gaan,li2018dcrnn,yu2018spatio}. 
However, such RNN-based temporal dependency modeling methods have a limitation in that they cannot effectively capture long-range temporal trends in a given sequence. 
To alleviate this weakness, recent models have employed other advanced techniques, such as convolution neural networks~\cite{Wu19} or self-attention networks~\cite{park2019stgrat}. 
In this work, we use \stgrat to model both spatial and temporal dependencies due to its enhanced interpretability and better performance over other models for traffic prediction~\cite{park2019stgrat}.
\subsection{Visual Analytics for Deep Learning Models}
Although deep learning models are effective, it is difficult to understand how the models work due to their black-box characteristic~\cite{Muhlbacher14}. 
Existing visual analytics approaches and systems for the matter fall into global model analysis and instance-based analysis~\cite{hohman2018visual}. 
The global model analysis systems focus on how to visualize internal model structures through a graph structure, whose \rev{nodes} and links are mapped to neural network neurons and weights between two connected neurons, respectively \rev{(e.g., CNNVis~\cite{liu16cnnvis})}.
In general, this global model analysis category includes visual analytics systems for understanding convolution neural networks (CNNs) applied to computer vision tasks~\cite{liu16cnnvis, pezzotti2018deepeyes,yosinski2015deepviz,revacnnfilm,zeiler2014visualizing} and recurrent neural networks (RNNs) for natural language processing tasks~\cite{ming2017understanding,strobelt2017lstmvis} 
For example, Liu et al.~\cite{liu16cnnvis} proposed a visual analytics system, called CNNVis to visualize a CNNs via a directed cyclic graph. 
They additionally use the edge bundling technique to visualize the learned filters and connections between layers. 
Zeng et al.~\cite{zeng2020revisiting} study a visual analytics approach to understanding the behaviors of CNNs, solving the modifiable areal unit problem.
\rev{Shen et al.~\cite{shen2020visual} suggest a visual analytics system that allows users to visually explore the model behavior from global and individual levels on a multi-dimensional time-series forecast.}
\rev{Compared to the work in the global model analysis category, the main objective of our work is different. }
For example, while their approach focuses on the analysis of input data with errors, aiming at stabilizing feature embedding and predictions, our work allows for the investigation of spatio-temporal dependencies that a model learns during training. 
Also, our automated approach is different in that it enables users to directly estimate the reasons for model behaviors using time-series causality analysis.

Fred et al.~\cite{hohman2019summit} present SUMMIT, which summarizes CNNs at scale by generating an attribute graph from a trained model and finding which filters influence on performance. 
Inspired by these studies, many VA systems have been proposed to solve similar problems with various models, including generative adversarial networks~\cite{kahng18GanLab,wang2018GANViz}, deep reinforcement learning~\cite{wang2019dqnvis}, and sequence-to-sequence models with attention~\cite{strobelt_seq2seq-vis_2019} and self-attention networks~\cite{cbpark2019sanvis,derose2020attention}.

The VA systems for instance-based analysis aim to show how an input instance influences a model and vice versa to help users identify models' robustness and internal mechanisms in given scenarios~\cite{Kahng2017activis,cabrera2019fairvis,Wexler20,liu2018analyzing}. 
For example, Strobelt et al.~\cite{strobelt_seq2seq-vis_2019} proposed Seq2Seq-VIS to help users manipulate models' internal parameters, and observe how models react to different inputs.
SANVis~\cite{park2019stgrat} visualizes self-attention networks with a given sentence, displaying how the model captures the linguistic relationships between the words in the sentence.

\subsection{Visual Analytics for Traffic Congestion}
As traffic congestion is a crucial problem; affecting life quality, many VA systems have been proposed to find the causes and possible solutions. 
Many computational and strategical approaches are employed in designing VA systems for traffic analysis.  
The systems often integrate new visualizations to better present traffic patterns and anomalies, such as T-Watcher with fingerprint visualization~\cite{Pu13}.
Wang et al.~\cite{Wang13} showcases an automated method for detecting traffic congestion, enabling users to explore different traffic congestion propagation graphs of a large city. 
Zeng et al.~\cite{Zeng13} proposed an interchange circus diagram to present interchange patterns at a junction road and help users identify multi-spatial scales and temporal change patterns. 
Recently, Pi et al.~\cite{Pi19} have built cumulative vehicle count curves (N-curve) and classified patterns of congestion cause by using entropy from information theory to analyze causes of congestion across intersections.

While much research has been conducted, few studies on visual tools exist for predicting traffic speeds and volumes, which is an important task for experts in the traffic domain, such as reporters at broadcasting centers. 
Lee et al.~\cite{Lee19} proposed a visualization system to help the experts whose tasks include broadcasting traffic conditions across a city. 
They use and train Long Short-Term Memory (LSTM) ~\cite{Hochreiter97} with three features--road network structure, neighbor roads' speed, and rush-hour--to improve prediction performance. 
However, as the system does not explain the prediction results, experts may not be confident when they broadcast possible congested roads to citizens based on the prediction results. 
In this work we design a visual tool that can help users understand the prediction results.

\section{Task Description and Requirements}
\label{task_analysis}
Our system has been designed with input from three domain experts of a corporate, which has 19 million users and processes 100 million map and navigation services per day, as of 2021. 
We have met with experts over a period of 18 months to extract task requirements and discuss design considerations.
The first expert (E1, 20 years of experience) is a director of the map service group, supervising all map-related developments and services.
The second expert (E2, 12 years of experience) leads the traffic intelligence team of the group and has developed algorithms for navigation services with the third expert (E3, 6 years of experience).
E2 and E3 are the machine learning and software engineers in the traffic domain, and both have software development skills and machine learning knowledge and have applied machine learning algorithms to the domain area (i.e., trainers~\cite{strobelt2017lstmvis}).

According to the experts, the group's main mission is to provide the best optimized navigation services to drivers and pedestrians. 
To fulfill this mission, they collect traffic speeds and paths from users (e.g., drivers who use their navigation services).
They also collaborate with traffic departments of the national government, police agencies, and city halls to collect and utilize additional road events, such as accidents and construction in their services.  
The collected data is used for pathfinding and travel time estimation algorithms.

One of the concerns of the group is that although they have done well in the domain, it is possible that their competitors can take a technical advantage with the recent rise of machine learning models, which mean a loss in their market share in the country.  
To overcome this, they have utilized DL technologies~\cite{yu2018spatio, Zhang2018gaan, Wu19, Pan2018HyperSTNetHF, Lee21}, including LSTM~\cite{Lee19}, attention-based models for predicting future traffic speed~\cite{park2019stgrat} and arrival time~\cite{derrow2021eta,fang2020constgat}. 
\rev{But, they often obtain unsatisfactory performance from the models, when they directly apply the models to their data. }
\rev{To find possible reasons for low performance, they conventionally investigate the model behaviors of roads with low accuracy, using the what-if method~\cite{Wexler20, whatif_pair, whatif_google}, in which they change the input traffic speed or features of a road in a model and compare forecasting performance. }
They also compare a road's traffic behavior to that of other roads. 
They know that a city has multiple regions, and each region shows different traffic and congestion patterns due to the various roles and capacities of the roads across the regions (e.g., residential and university areas, industrial complexes~\cite{Lee21}). 
In terms of a navigation service point of view, finding out which time and locations where model produces high accuracy is important information for deployment.

But, after starting their investigation, they soon find that this case-based, individual prediction performance analysis with Python is onerous and ineffective.
In particular, the approach takes a great deal of time and effort to answer all of the questions that they create at work for achieving their goal and deployment criteria due to a large number of road links in cities~\cite{Lee19} and complex dependencies among roads~\cite{guo2019identifying}.

The questions that need to be answered for their tasks are as follows: \textbf{(Q1) Where and when does a model produce inaccurate predictions?; (Q2) What could be the causes of inaccurate predictions?; (Q3) Which past (combinations of) observations does a model use for speed prediction?; (Q4) Which road does a model refer to for speed prediction?; (Q5) How can the accuracy of a model be improved?; and (Q6) How does a model work across different road topologies?}
Among the questions, Q1 and Q2 are the main questions, while Q3 and Q4 are auxiliary questions that need to be clarified to better answer Q1 and Q2. 
Q5 can be considered a question for increasing their competence in the market and Q6 is the last question that should be addressed before in-the-wild deployment.

From the discussions, we have derived the following requirements for a visual analytics system designed for exploring traffic forecasting model's behavior and performance improvements.
First, as it is important to pop out problematic roads with a given model, \textbf{(R1) a VA system should highlight the roads with low accuracy and provide information on when a model has low accuracy (Q1)}. 
As models predict future speed based on spatio-temporal dependencies among roads, \textbf{(R2) a system should provide a method for effectively exploring encoded dependencies (Q2--Q4)} so that users can find evidence on the relationship between high errors and speed patterns~\cite{chen2012retrieval, li2018brief}.
Example information for supporting spatio-temporal dependency exploration includes \textbf{(R2-1) historical traffic patterns of roads, speed distribution of data, standard deviation, daily speed trends (Q3, Q4)}, and \textbf{(R2-2) data on model behaviors (Q2--Q5)}, such as which roads a model refers to for forecasting  (i.e., which roads influence the prediction for target road~\cite{guo2019identifying}), and which input sequences are importantly used for the prediction. 
In many cases, it is difficult for users to identify which roads affect the prediction performance of a certain road~\cite{li2018brief}. 
While prior work has revealed~\cite{guo2019identifying} that there is a strong dependency among neighboring roads and second and third linked roads by cross validation, methods for investigating whether a target road refers to appropriate neighboring roads or not have been rare. 
Therefore, \textbf{system should also provide information on the similarity of temporal data and causality among roads (Q2-Q4).}
Lastly, to support users' formulation and validation of hypotheses, \textbf{(R3) a system should provide a method that shows how much improvement users can expect (Q5, Q6).}

\section{Data Description and Attention Mechanism}
\label{sec_stgrat}

\begin{figure*}[t]
  \centering \includegraphics[width=0.8\textwidth]{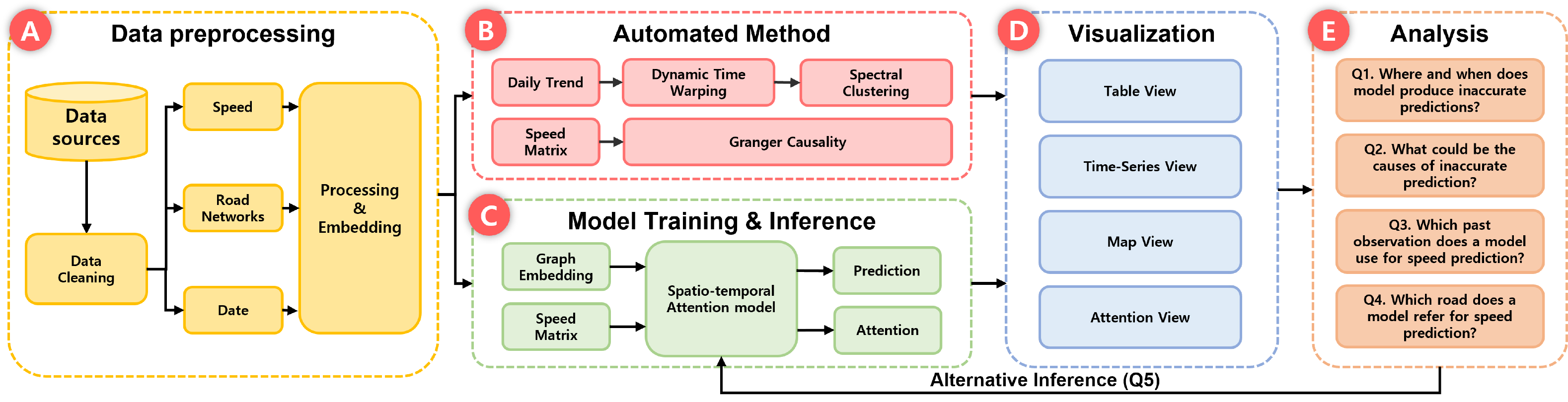}
  \vspace{-0.3cm}
    \caption{An overview of the system workflow with functional modules and questions. (A) Data pre-processing, (B) automated methods for supporting spatio-temporal analysis, (C) model training and inference, (D) visualization modules, and (E) answering questions using the system. 
    }
    \label{fig_intro}
    \vspace{-0.6cm}
\end{figure*}

\begin{figure}[t]
  \centering \includegraphics[width=0.8\columnwidth]{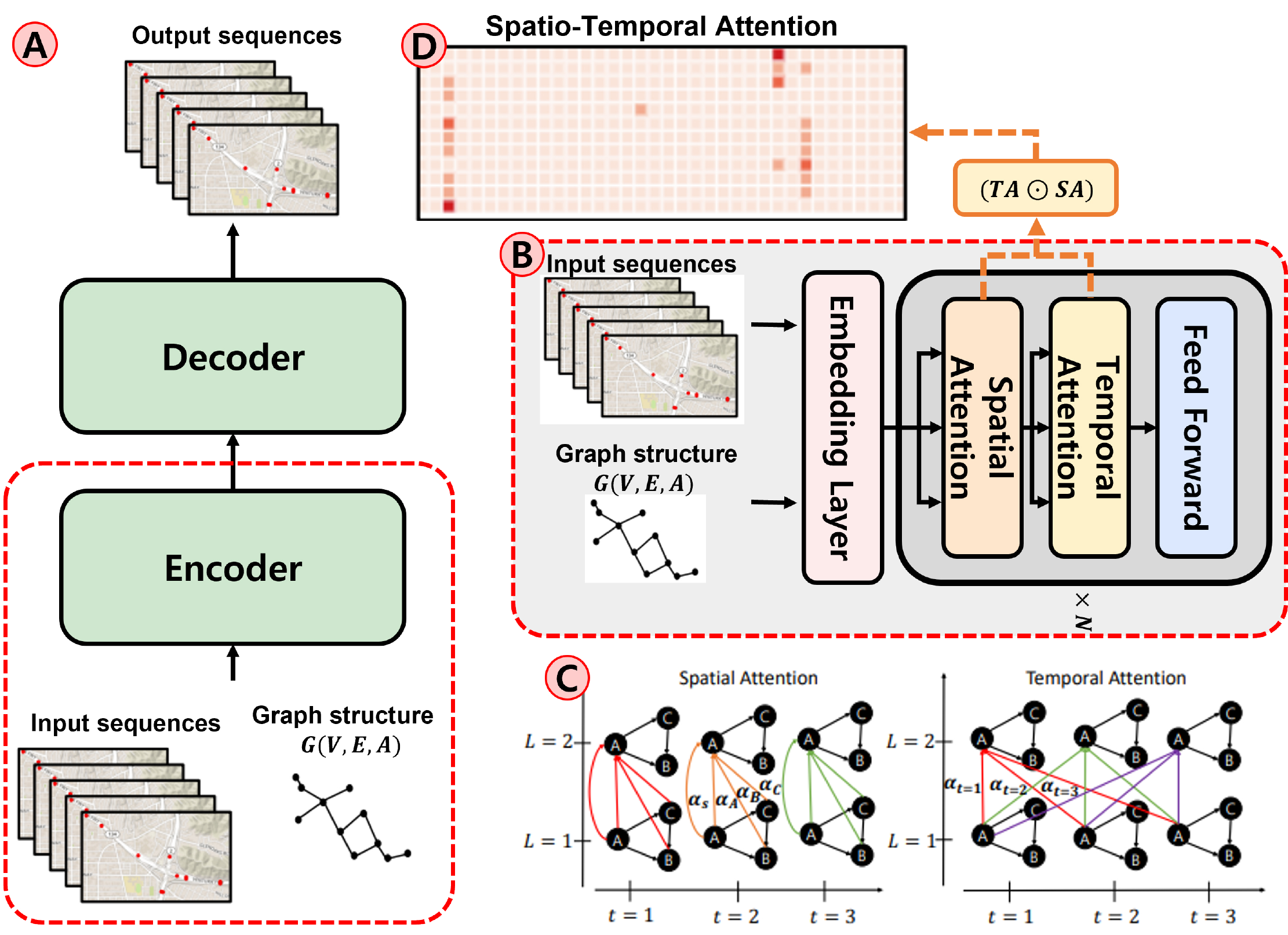}
    \vspace{-0.3cm}
    \caption{Overall architecture of \stgrat (A). Each layer in the encoder is composed of a stack of identical layers (B). G($V$, $E$ and $A$) in (A) indicates a set of nodes, links, and a weighted adjacency matrix, respectively. We create a spatio-temporal attention matrix (D) by calculating point-wise multiplication between spatial and temporal attention weights \rev{(C)}.
    }
    \label{fig_stgrat_overview}
    \vspace{-0.6cm}
\end{figure}

\subsection{Data Description}
\label{sec_data}
In this work, we use the traffic data of two different road networks--the urban and highway road networks to explore the model's inference process for speed prediction. 
For the urban road network, we use dedicated short range communication (DSRC) data~\cite{Kenney11} generated from Ulsan, South Korea (range: 9/1/2017$\sim$12/28/2017), where more than 1.1 million people live with more than 540,000 registered vehicles as of 2017. 
A total of 116 DSRC sensors are used for data collection, which are installed every 5.7km and cover a total of 68 main roads. 
For the highway road network, we use the METR-LA data~\cite{Jagadish14}, which were collected from 207 loop detectors (range: 3/1/2012$\sim$6/27/2012) on the highways of Los Angeles.
Note that the highway network data we use are the standard benchmark data for traffic forecasting tasks~\cite{li2018dcrnn,yu2018spatio, Zhang2018gaan}.
\rev{After discussing with domain experts and reviewing training results, we replace the missing data and explicit errors with historical data. 
We also use 5-minute aggregated data to mitigate possible effects of outliers, as performed in many previous studies (e.g., \cite{li2018dcrnn}).}

\subsection{Attention Mechanism}
We describe how we utilize \stgrat to demonstrate our VA approach. 
We choose \stgrat because 1) it has demonstrated state-of-the-art performance and because 2) we can produce attention matrices on spatio-temporal dependency (e.g., Eq.~\ref{eq_attn}, Fig.~\ref{fig_stgrat_overview}D)~\cite{park2019stgrat}.
We also consider that the domain experts have expressed that attention models are extensively used at work for not only speed prediction but also other tasks, such as travel time~\cite{fang2020constgat} and taxi demand prediction~\cite{yao2019revisiting}.

\stgrat (Fig.~\ref{fig_stgrat_overview}) is a variation of the transformer~\cite{vaswani2017attention} that uses the encoder-decoder architecture with self-attention (i.e., temporal attention). Additionally, \stgrat utilizes graph attention as spatial attention before temporal attention with a sentinel vector. 
\rev{The sentinel vector acts as weights for skip connection within the same road.}
\stgrat utilizes 12-length sequential historical speed with encoded features for each road and predicts 12 sequential speed predictions.

There are three types of layers in the encoder and decoder: embedding, spatial attention, and temporal attention layers (Fig.~\ref{fig_stgrat_overview}B). 
To allow the model to extract the \rev{spatio-temporal} dependencies, we provide a road network, speed, and observed time as the input features for the embedding layer (Fig.~\ref{fig_stgrat_overview}B). 
The road network graph is directed graph $G(V, E, A)$, where a road is represented as a node (i.e., $V$) and the connection between roads is shown as a link (i.e., $E$). 
\rev{Note that road network $G$ is directed graph, which allow model to distinguish the road directions.} 
Note that we provide the order of a given sequence using the position embedding method~\cite{vaswani2017attention}.

The model captures spatial dependencies among neighbor roads in the spatial attention layer (Fig.~\ref{fig_stgrat_overview}B) by using a graph attention network~\cite{Velickovic2018graph, Zhang2018gaan}, a well-known graph modeling method. 
In short, the spatial attention layer integrates information among neighboring roads by directed graph attention. 
This directed spatial attention improves dependency modeling and helps developers interpret \stgrat.

The temporal attention layer (Fig.~\ref{fig_stgrat_overview}B) models the temporal dependency and trends of given sequences. 
For modeling, temporal attention performs multi-head attention to compute temporal correlations. 
\rev{The attention type is decided by attention axes; spatial attention aggregates features among the spatial axis (i.e., neighboring roads), while temporal attention aggregates input features within the temporal axis (i.e., different time steps of the same roads (Fig.~\ref{fig_stgrat_overview}C).}

To sum up, there are two important attention layers--spatial and temporal attention--in \stgrat. 
Vaswani et al.~\cite{vaswani2017attention} describe an attention function as mapping a query and a set of key-value pairs to an output, where the query, key, values, and output are all vectors. 
From this perspective, from each key-value pair, we can derive an attention matrix from each attention layer, called \textbf{SA} and \textbf{TA}, and create a \textbf{spatio-temporal matrix (ST matrix)}. 
Specifically, for each attention head, given $X$ before spatial attention and $H$ after temporal attention, their relationship can be written as follows:
\begin{align}
\label{eq_attn}
    H &= (\textbf{TA} \odot \textbf{SA}) X,
\end{align}
where \textbf{TA} $\odot$ \textbf{SA}, named \textit{spatio-temporal attention}, is the attention that users mainly explore for understanding model behaviors.

\section{Visual Analytics Environment}

We describe our visual analytics system, \toolname, which we have designed to effectively explore the deep learning model's internal process for traffic forecasting using attentions. Fig.~\ref{fig_intro} shows the pipeline of the analysis with \toolname. 
We first prepare the data (A) for automated methods (B) and model training (C). 
After model training, we extract ST attention from the model (C), which is used for visualizations (D). 
Users explore model behaviors using the automated methods, (E) answering the questions (Q1--Q4) and test and confirm performance improvements (Q5).

\subsection{Automated Methods}
\label{sec_algorithm}
In this work, we incorporate two automated methods--\textbf{dynamic time warping (DTW)} with spectral clustering~\cite{salvador2007toward,berndt1994using, xiao2016traffic} and \textbf{Granger causality tests}~\cite{diks2006new}, to allow users to determine which references are appropriate for predictions and whether a model employs proper references for the predictions (R2-3)~\cite{li2015trend}.

We use daily trends of roads as input for DTW computation, which should be explored since they are encoded as temporal dependencies among roads, which a model learns during training~\cite{yao2019revisiting}.
However, it is difficult to analyze road trends for two reasons. 
First, there are too many trends (i.e., roads) in a traffic dataset. 
Second, there are different time gaps between the trends of roads due to different levels of dependencies, which are presented with either lagging or preceding trends~\cite{li2018brief}. 
For example, when congestion occurs on a road, the roads linked to the congested road also become congested, but there are no concrete patterns when the neighboring roads are congested. 
As such, we incorporate dynamic time warping (DTW)~\cite{berndt1994using} in this work, which calculates the similarity in two time series over time gaps after finding the best matching alignment that minimizes the distance, as Le Guen and Thome do for their time-series forecasting model design~\cite{le2019shape}.

Then we perform spectral clustering~\cite{miyahara2014family} with the computed DTW scores to further allow cluster-based analysis, which has shown its effectiveness in analyzing traffic data~\cite{aven2017daily, xiao2016traffic}. 
For example, users can easily confirm whether referred roads of a target road have similar daily trends as the target road using clusters. 
If the target and referred roads are in the same cluster, it is highly possible that the model learns the dependencies of the target and referred roads during training~\cite{li2018brief}.
Note that we use the ELBOW method~\cite{afzalan2019automated} with visual inspection to choose the number of clusters.

As Wu et al.~\cite{wu2020connecting} show in their study, catching preceding patterns in time-series enables effective analysis of deep learning models.
Therefore, to help users better explore the preceding patterns in the traffic data, \rev{and to complement DTW that distorts the time during its computation,} we incorporate the Granger causality test~\cite{diks2006new} in this work, a well-known temporal dependency investigation method.
We first describe the definition of Granger causality based on two principles: 
\textbf{(1) the cause happens prior to its effect and (2) has unique information about the future values of its effect.}
Given these two principles about causality, Granger causality proposes testing the following hypothesis for the identification of a causal effect.
\[ Granger Causality = \mathbb{P}[Y(t+1) \in A \vert \mathcal{I}(t)] \neq \mathbb{P}[Y(t+1) \in \vert \mathcal{I}_{-X}(t)] \]
Here, P is probability, A is an arbitrary non-empty set, and $\mathcal{I}(t)$ and $\mathcal{I}_{-X}(t)$  denote the information available as of time t in the entire universe, and in the modified universe where X is excluded, respectively. 
If the above hypothesis is accepted, we say X Granger-causes Y.

\subsection{Visual Interfaces}
\label{sec_interface}
\toolname consists of \rev{five} views: (A) filter, (B) table, (C) line, (D) map, and (E) attention, as shown in Fig.~\ref{fig_teaser}. 
The attention view (E) has two sub views--spatio-temporal and head cluster (Fig.~\ref{fig_headcluster}) views. 
In the filter view (A), users can select a dataset, date range, and target prediction time (e.g., 15, 30, 45, and 60 minutes; default: 15 minutes).

Users can filter out the roads using Mean Absolute Error (A1:\iconMAE: scaled: min--max values) and spatio-temporal attention (A2:\iconAttn scaled: 0--1) values.
When roads' MAEs are higher than the threshold value, the center circle of the roads are colored in black \textbf{(R1)}.
If a road's MAE value is below the value at Q1 (i.e., top 25\% in accuracy), the road is included in the low error group.
If the MAE of a road is above the value at Q3 (i.e., bottom 25\% in accuracy), it belongs to the high error group. 
When users hover over the filters, a tooltip pops up to show the MAE values at the Q1 and Q3 boundaries (Fig.~\ref{fig_teaser}, A1). 
The attention filter is applied to \ars (e.g., Fig.~\ref{fig_arrow}) on the map, when a road is selected for investigation. 
There are two legends to represent clusters (Fig.~\ref{fig_teaser} D1) and ratio values (D2).
We use the color sets from ColorBrewer~\cite{harrower2003colorbrewer}.

\subsubsection{Table View}

The sortable view (Fig.~\ref{fig_teaser} B) allows users to explore roads' characteristics in a compact form with information on road ID (Road), speed distribution (Speed Dist. (std.)), speed trends (Trend), cluster index (CLS), mean absolute error (MAE), and causality analysis (CA).
The view is initially sorted by clusters, but users can sort differently using a small triangle (\iconTriangle) at column titles.
The column title is underscored to represent which column is used for current sorting in the table. 
For example, the table is currently sorted by the standard deviation (i.e., std) of roads' speed (Fig.~\ref{fig_teaser}, Speed Dist. (std)).
We show the speed distribution of a road (Speed Dist.) in a histogram with the x-axis representing different speed ranges with bins (e.g., 10 miles/bin) and the y-axis meaning the frequency.
The bins' height is normalized by the maximum count. 
At the right side of the distribution, we also place each road's standard deviation value. 
If a road has a high deviation score, it means the road speed often significantly changes, while the low deviation implies that significant speed changes rarely happen on the road (e.g., highways).

To allow speed trend analysis, we provide normalized speed trends using spark lines~\cite{tufte2006beautiful} (Fig.~\ref{fig_teaser}B, Trend column), with the x-axis representing the time range (0--23) and the y-axis denoting average speeds of roads with the 5-minute interval.
As users hover over a trend line, a vertical line and tooltip appear, showing time stamps and speeds.
After the trend column, we provide cluster indices that each road belongs to and the MAE values of roads to help users investigate the relationship between speed trends, road clusters, and model accuracy \textbf{(Q2, Q3)}.
When users click on a trend line, a new window pops up, and shows speeds and MAE values.

We present Granger causality test results (F values) in the last column (CA) to support users inspecting spatio-temporal dependency in terms of causality.
When users click on a road on the table, the F-values between the clicked road and other roads are presented.
If the F-value of a pair is higher than of other pairs, we can say that the pair with a higher F-value has a more significant Granger causality relationship than other pairs. 
For example, we can see from Fig.~\ref{fig_ulsancase} B1 that the Road 113 and Road 112 pair has a more significant Granger causality relationship than Road 113 with other roads. 
Note that if the p-value of a pair is less than 0.05, we do not show the results.
We provide the test results to help users make and validate hypotheses on model behaviors.
For example, users can review F-values to determine if the model sufficiently refers the information of the road with a significant Granger causality relationship for performance analysis \textbf{(Q2, Q5)}.

Users can perform interactions in the table view for coordinating multiple views for detailed investigation. 
When users hover over a road in the table, corresponding road on the map and the attention matrix are highlighted.
When users click a road id, (1) table view highlights (light-blue) the row of the road (e.g., Fig.~\ref{fig_teaser} B2), (2) map view places the road at the center of the map with highlighting in light blue, and (3) line view shows the collected and predicted speed lines (Fig.~\ref{fig_teaser} C).

\subsubsection{Time-Series Views}
To help users investigate roads' temporal dependencies \textbf{(R2)}, comparing road speeds, we provide a line chart view (Fig.~\ref{fig_teaser}C), where the x-axis and y-axis represent time (5-minute intervals) and road's speed, respectively.
Users can add raw and predicted speeds by selecting roads from other views, such as table, map, and attention views, and remove the lines by toggling road labels at the top.
There is a blue vertical bar in the view (Fig.~\ref{fig_teaser} C1), indicating the current data point in visualizations (e.g., ``2017-12-17 13:35:00," Fig.~\ref{fig_teaser}C, top blue panel).
It also shows the absolute error (AE) and STD values in the past one hour of the time specified by the bar (e.g., ``AE: 1.24 STD:160.70" Fig.~\ref{fig_teaser} C1). 
Users can move the bar to anywhere to update map and attention views with the data specified by the bar.

\begin{figure}[t]
    \centering 
    \includegraphics[width=0.6\columnwidth]{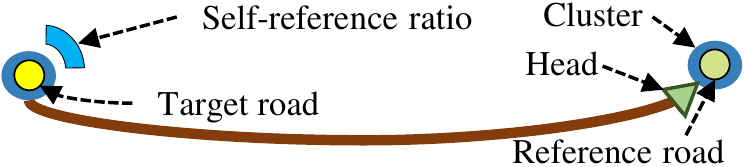}
    \vspace{-0.3cm}
    \caption{\ar for presenting which roads a target road refers to for inference. The donut chart surrounding the target road is filled in blue (clockwise) based on the self-reference intensity.}
    \label{fig_arrow}
    \vspace{-0.6cm}
\end{figure}

\subsubsection{Map View}
\label{map_view}

As models learn dependencies among roads and forecast based on the learned dependencies, it is important to explore what dependencies a model learns with road speed patterns and which roads a model refers to for performance investigation.
To help users perform such exploration \textbf{(R2-1, R2-2)}, we first present each road as a white dot.
Each circle surrounding the dots shows the cluster that the road belongs to. 
We provide three visualizations in the map view: heatmap, attention arrows (\ars), and cluster visualization. 
First, to help users overview the relationship between roads' congestion levels and model performance \textbf{(Q1)}, we visualize roads' congestion levels using heatmaps (Fig.~\ref{fig_teaser}D). 
Here, the redder the heatmaps are, the slower the roads are (heatmap legend: Fig.~\ref{fig_teaser} D2).
We also have considered providing heatmaps with a time filter so that users can explore regions with high model error but decided against this because this approach requires many interactions for filtering and memorizing heatmaps that have changed due to filtering.
Using Bezier curves, \textbf{\ar} (Fig.~\ref{fig_arrow}) link a target road and the roads that the target road attends for prediction.
Here, the color represents the amount of attention given to the roads.
The darker the head color, the more attention the reference road is given. 
\rev{For example, the head near Fig.~\ref{fig_teaser}, 73 shows that the attention level is about 40\% of the maximum attention value (legend: Fig.~\ref{fig_teaser} D2).}
To reduce visual clutter, users can hide unimportant \ars by using attention filters (Fig.~\ref{fig_teaser} A2), or click \ars to turn off~\cite{ko2014analyzing}. 
By default, the \ar head visualizes the encoder's attention data.

When users click a road to set a target road for investigating its reference roads, the dot is highlighted steelblue (e.g., Fig.~\ref{fig_teaser}D, road 81). 
To represent the selected target road's self-reference intensity (i.e., how much a road attends to its own speed pattern), we present a donut chart around the road, filling the chart according to the self-reference intensity  (scale: 0--1, clockwise).
For example, the donut chart of road 81 (Fig.~\ref{fig_teaser}D) has self-reference intensity of 0.29. 
There are buttons for interactions on the map (Fig.~\ref{fig_teaser} D3)--CA (Causality Analysis), CLR (Clear map), LBL (Label display), JAM (Traffic Jam Heatmap), zoom-in, and zoom-out.

When users click the CA button, \toolname turns on the Causality Analysis mode.
Then if users click a road, \toolname removes the roads on the map, whose p-value in the test is higher than 0.05 to allow users to analyze with statistically meaningful roads.
Users can draw a selection area by creating a polygon on the map where a set of roads can be included for investigation and un-select the selected roads using CLR. 
The speed heatmap and road labels are turned on and off with JAM and LBL, respectively.

\begin{figure*}[ht]
  \centering \includegraphics[width=1\textwidth]{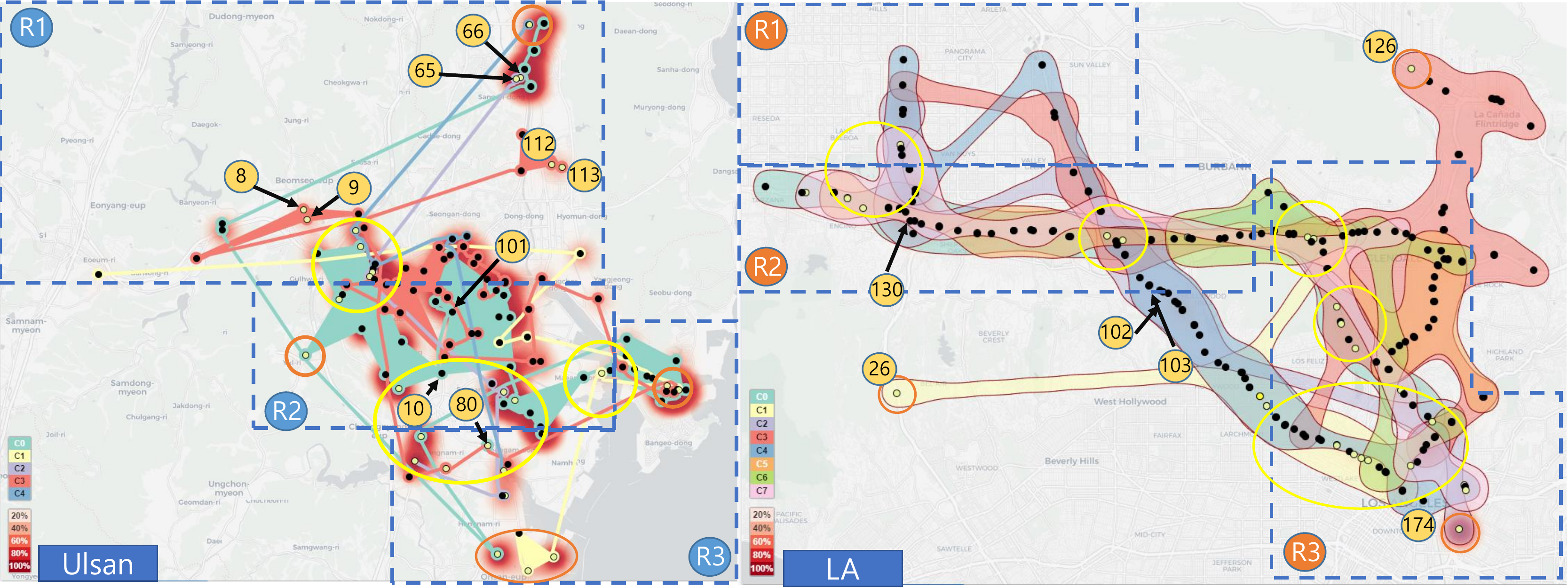}
    \caption{We both implemented KelpFusion and BubbleSet in the system for the cluster visualization and found that KelpFusion works well with a complex road network (left), while BubbleSet effectively shows clusters by connecting highway roads in linear shapes. However, it becomes harder to identify clusters with the two visualization. Thus, we propose a circle-based cluster visualization in this work (\autoref{map_view}).}
    \label{fig_data}
\end{figure*}

To help users find which road belongs to which cluster, we initially implemented and compared BubbleSet~\cite{collins2009bubble} and Kelpfusion~\cite{meulemans2013kelpfusion}, which are effective for set visualization.  
From our observations, although effective with a small number of clusters, they produced several overlaps among clusters when roads in different clusters are located closely. 
We have also observed that they may not be helpful, as a large number of overlaps means that there rarely is a dominant clustering pattern among the neighboring roads. 
Appendix Figure~\ref{fig_data} shows our implementation results with 5 and 8 clusters. 
After the investigation, we decide to encode the clusters by placing an outer, colored circle on each road~\cite{kwon18clustervision}.
Figure~\ref{fig_arrow}  shows two example roads, where both the target (left) and reference (right) roads belong to cluster C0 (legend: Fig.~\ref{fig_teaser} D1).

\subsubsection{Attention View}
\label{sec_attn}
There are two visualization tabs in the attention views: spatio-temporal (ST) and head-cluster views for \textbf{R2-2}. 
We use pixel-based visualization~\cite{ko2012marketanalyzer, keim2000designing} to present the model's spatio-temporal attention in a matrix form, because of its scalability on the number of items~\cite{keim2000designing, ko2012marketanalyzer}.
In the ST attention matrix view (e.g., Fig.~\ref{fig_teaser}E), the x-axis indicates different roads, and the y-axis indicates past time steps (5-minute intervals, 12 steps, top: 5 minutes ago, bottom: 60 minutes ago).
In the view, users can analyze which roads and time steps the model focuses on with attention intensity, represented by the color (legend: Fig.~\ref{fig_teaser} D2). 
For example, when users click road 81 on the map, the spatio-temporal view (Fig.~\ref{fig_teaser}E) shows that road 81 attends itself (i.e., self-reference) and 73 more than other roads for its prediction.

\begin{figure}[t]
  \centering
  \includegraphics[width=0.8\columnwidth]{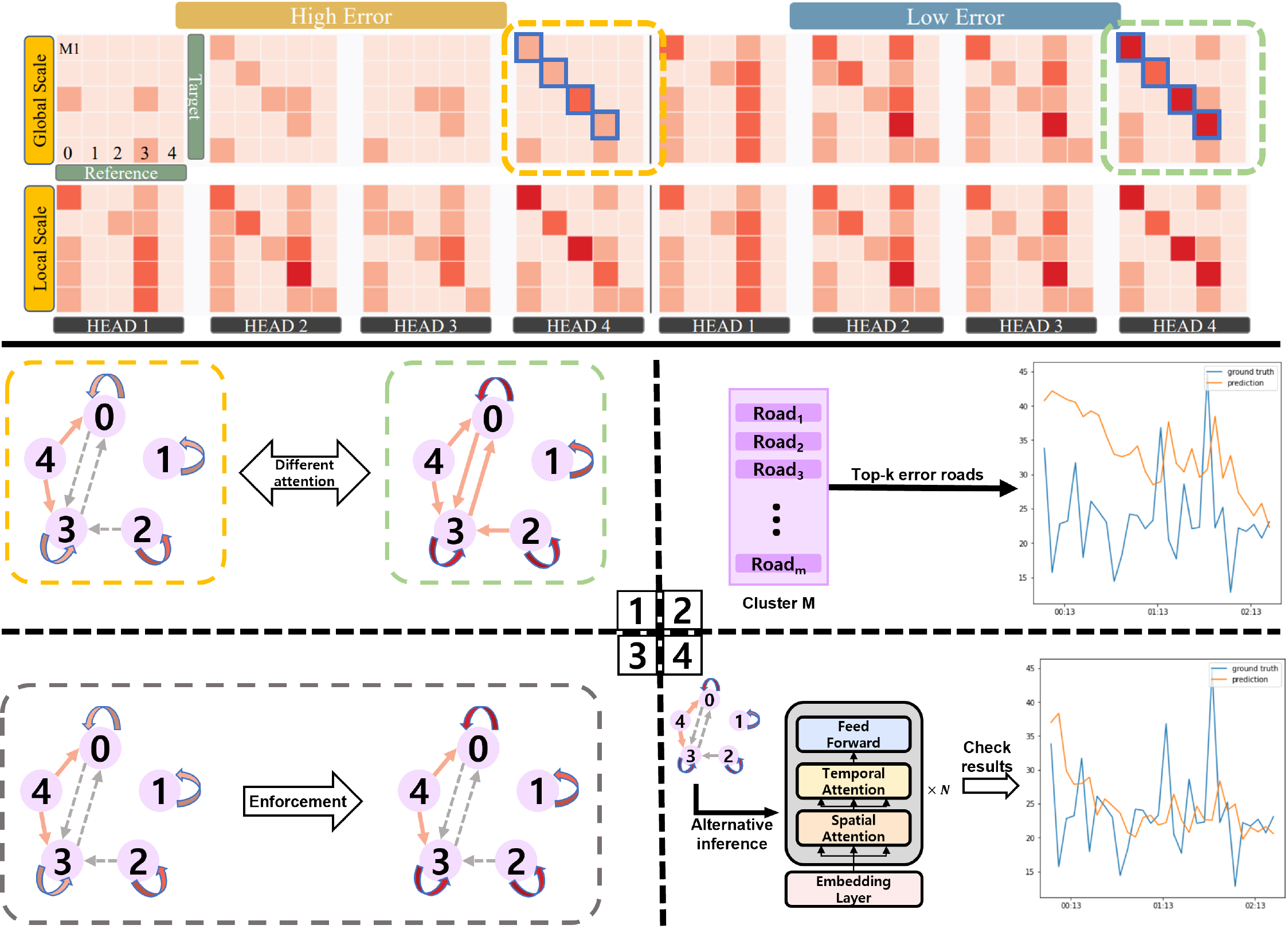}
  \vspace{-0.4cm}
  \caption{(Top) The head-cluster view with four attention heads, (Bottom) \rev{The enforcement process--1) comparing attention between low (left) and high (right) error cases, 2) selecting $k$ highest error roads in each cluster, 3) replacing the attention of the selected roads with that from low error roads, and 4) testing alternatives. }
  }
  \label{fig_headcluster}
  \vspace{-0.7cm}
\end{figure}

There are 8 matrices in the head-cluster view (e.g., Fig.~\ref{fig_headcluster} top, first row) to show how much each attention head refers to reference roads for making predictions for target roads in a cluster point of view. 
In the view, the first four matrices visualize the attention patterns of the four heads with the roads for which the model records high error (top 25\%), while the other four matrices display those of the four heads with the roads for which the model shows low error (lower 25\%).

In each matrix, the column represents the clusters of reference roads, while the row means clusters of target roads in ascending order.
The color of each cluster cell represents the intensity of the attention that each head assigns, so if a cell color is the darkest red, it means an attention head heavily refers to the roads in the cell (i.e., cluster) for making predictions (legend: Fig.~\ref{fig_teaser} D2).
For example, many target clusters attend cluster 0 and 3 (Fig.~\ref{fig_headcluster} top, low error) and record low error. 
This indicates that the roads in cluster 0 and 3 have strong similarity with the roads in the target clusters in terms of daily speed trend.

There are global and local scales to normalize the attention values (i.e., intensity) in the matrices differently. 
All attention values across matrices are divided by the largest attention value to show relative attention intensity in the global normalization (Fig.~\ref{fig_headcluster} top, first row), but in the local normalization, each cell is divided by the sum of each row to make row sums of the matrices equal to 1.
We show the attention patterns with the two scales to help users analyze the attention patterns by individual heads and across heads \textbf{(R2-2)} and find the reason for the failed inference, using the level of attention intensity information.

When users hover over a cell on a matrix, a tooltip pops up, showing detailed information, such as head index, relative attention ratio, and the average of their attentions.
The interaction also highlights the road in the map and table views associated with the hovered item.

Inspired by Strobelt et al.'s approach that nullifies attention~\cite{strobelt_seq2seq-vis_2019}, we present an \textbf{attention enforcement method} in the head cluster view that replaces the attention values of the roads with high error with those from the roads with low error. 
Fig.~\ref{fig_headcluster} shows the process. 
For example, as users select four clusters (blue outline) in the view, \toolname searches top-k target roads with high error in each cluster (Fig.~\ref{fig_headcluster} [2]). 
Then, the system automatically finds the most appropriate reference roads for each selected road in the chosen clusters, using the DTW distance matrix and Granger causality tests with equal importance (0.5).
Lastly, the system extracts the attention values of the reference roads (i.e., alternative inference, Fig.~\ref{fig_headcluster} [3]), applies alternative inference with the replaced attention to the target roads, and expresses new prediction results (Fig.~\ref{fig_headcluster} [4]).
Fig.~\ref{fig_headcluster} [1] and [3] show the differences in attention distribution among clusters (i.e., color darkness of the arrows) in the initial and final cases. 
When users click on the ``Test Alternatives'' in the view (Fig.~\ref{fig_teaser} E1), a new view pops up, showing two line charts to describe the model's original and resulting accuracy (e.g., Fig.~\ref{fig_report}).
Here, the x-axis corresponds to the degree of error, while the y-axis presents the frequency of the roads with specified error. 
In sum, if the original performance graph is moved to the left, there is an improvement in model performance. Detailed analysis with the result of using the method is presented in Sec.~\ref{sec_enforce}.

\section{Case Study}
\begin{figure*}[t]
  \centering 
  \includegraphics[width=0.85\textwidth]{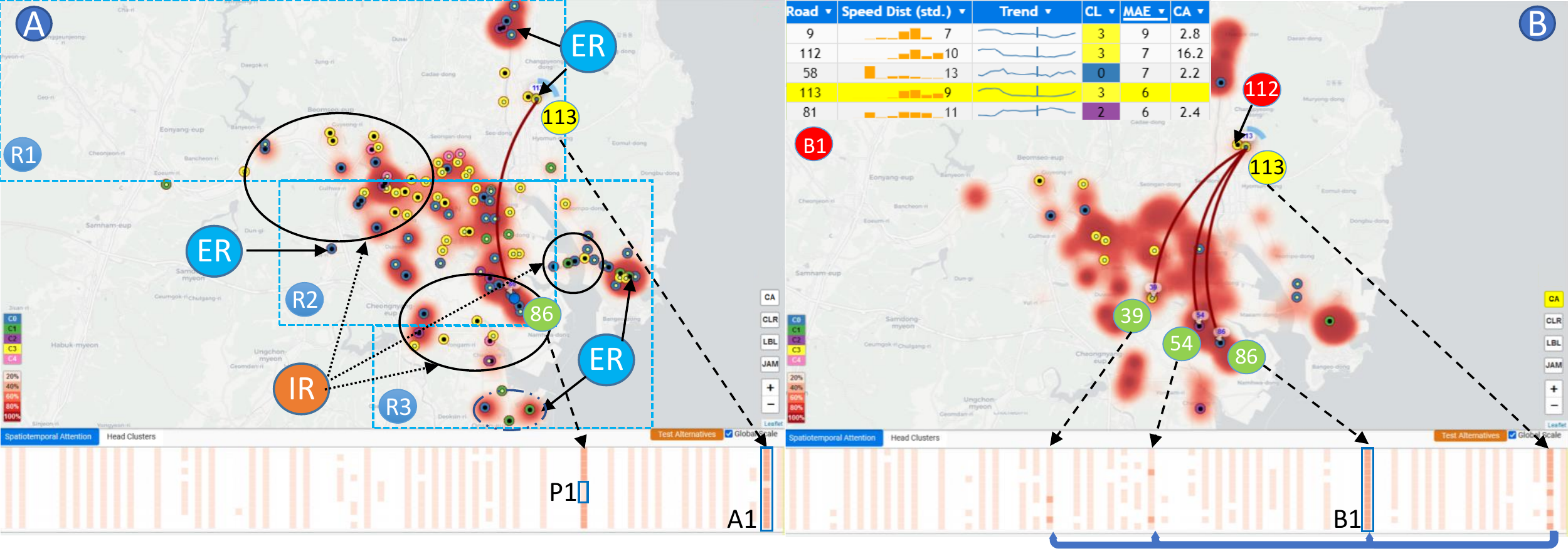}
  \vspace{-0.3cm}
  \caption{
  Two example visualizations during analysis. (A) There are three regions--R1: residential, R2: downtown, and R3: industry complex. R2 has heavy congestion during rush-hours due to in-and-out traffic from R1 and R3. 
  High error roads show inner black dots as a filtering result (Sec.~\ref{sec_interface}). 
  (B)Roads with low causality values and attention for Road 113 are filtered out in the causality mode. %Notice that for road 113, there are many roads which does have not attentions than filter (0.1) even in the same cluster including neighboring road 112.
  }
  %Difference of attention between the low and high error case in the urban road, Ulsan city, South Korea. There are three different regions, where R1 is the residential zone, R2 is downtown and residential complex, and R3 is the industrial zone. Traffic flow is R1 $>$ R2 $>$ R3 at the morning rush hours. Whitely highlighted dots in A, represent roads where high MAE (above Q3, 4.2).
  \label{fig_ulsancase}
  \vspace{-0.6cm}
\end{figure*}

We present two case studies with two different road networks--the urban and highway road networks--exploring the model's speed prediction process. 
The case studies are archetypal use-cases that drove the design along with the feedback from our domain experts. 
The third case study shows how model performance can be improved based on the findings derived from the two case studies. 
As described in Sec.~\ref{sec_data}, we use DSRC data generated from Ulsan, South Korea as urban road networks and METR-LA data as highway road networks.
Note that we define high error (low performance) and low error (high performance) groups of roads. 
The high error group includes the roads whose MAE is higher than the third MAE quartile (Q3), while the low error group include roads whose MAE is lower than the first MAE quartile (Q1).

We set DTW's window size as 4 to capture up to 20--minute time lags and use 5 and 6 clusters for Ulsan and LA, respectively, based on the elbow methods~\cite{pensky2019spectral}.
We performed pre-processing so that both data sets have a 5--minute interval of speeds and timestamps and replace missing values with averages of past data. 
We used 70\% of data for training, 10\% of the data for validation, and 20\% of the data for testing.

\begin{figure}[t]
  \centering
  \includegraphics[width=0.85\columnwidth]{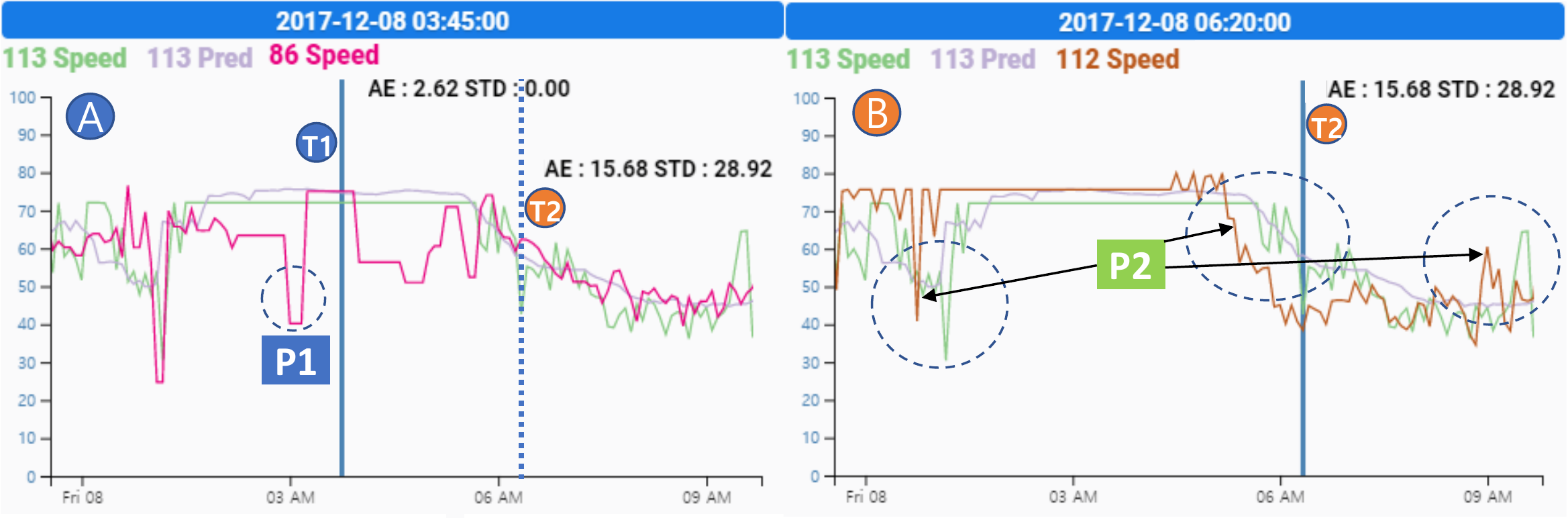}
    \vspace{-0.3cm}
  \caption{Inaccuracy is related to a high standard deviation of the speed trend in a given window. The model records a low error at T1 (low deviation) and a high error at T2 (high deviation). The traffic pattern of Road 112 precedes that of Road 113 (P2).
  }
  \label{fig_ulsan_sptn}
  \vspace{-0.7cm}
\end{figure}

\begin{figure*}[t]
  \centering
  \includegraphics[width=0.85\textwidth]{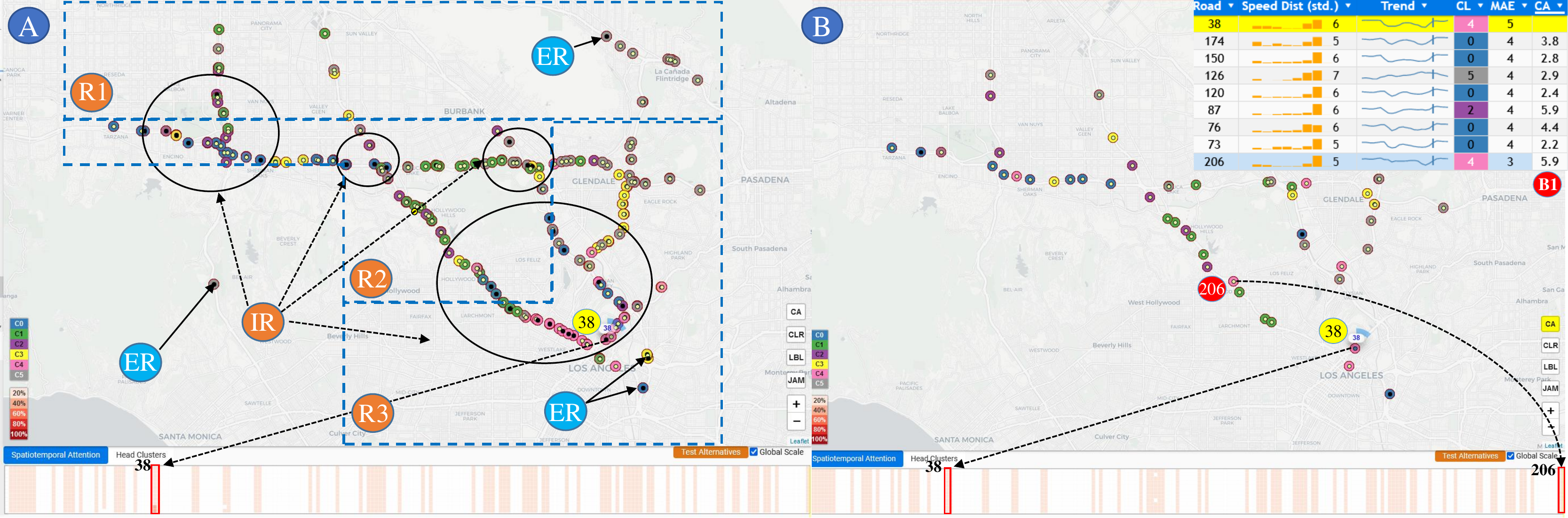}
    \vspace{-0.4cm}
  \caption{An example visualization view with the highway speed data in L.A. (R1: residential, R2: downtown, R3 industrial regions). The two attention views show how the model attends in the low (A) and high error (B) cases. 
  }
  \label{fig_lacase}
  \vspace{-0.5cm}
\end{figure*}

\subsection{Analysis with Urban Roads}

A machine learning model developer, Jane, has recently developed a new deep learning model for traffic forecasting using the attention mechanism and found that the new model shows higher accuracy in experiments, compared to existing models (Appendix, Table~\ref{table_result_metr}); however, she believed the accuracy can be improved more, if she understands how the model works. 
So, she decided to use \toolname to explore how the model predicts road speeds.

After the traffic data of Ulsan was loaded, \toolname showed the roads on the map and the detailed information on the roads in the table (Fig.~\ref{fig_teaser}B).
Jane then turned on the speed heatmap to obtain an overview of typical traffic congestion across the city by clusters and to analyze error patterns by cluster.
For example, she found that Cluster 3 usually experienced traffic congestion during the morning and evening rush hours as it \rev{includes} many roads in the downtown region (Fig.~\ref{fig_ulsancase}A, R2), while Cluster 4, which is in a residential region (Fig.~\ref{fig_ulsancase}A, R1) is congested in early morning~\cite{Lee19}.
Next, as she set the MAE filter's threshold at 4.2, which is about one--fourth of 11 (the largest MAE, Fig.~\ref{fig_teaser} A1), \toolname highlighted the roads with low accuracy on the map.
Interestingly, there seems to be no strong correlation between traffic congestion and model performance.
Instead, it was shown that high error roads tend to be located at edges of regions (ER) and intersections of regions (IR), such as commercial and industry complexes, and residential regions~\cite{Lee19} as shown in Fig.~\ref{fig_ulsancase}A.

As the roads with similar characteristics may have similar error levels, she found it is important to explore the roads of high errors \textbf{(Q1)}.
To identify the characteristics in common, she sorted the table, making hypotheses on a relationship among the deviation, distribution, and accuracy. 
She quickly noticed many cases seem to imply a relation between speed distribution patterns and the road error, MAE in the table view (Fig.~\ref{fig_teaser}B).
For example, she found that road 81 and road 66 (Fig.~\ref{fig_teaser} B3) had similar speed distribution, deviation, and error. 
They both had high deviation (e.g., 11, the speeds were similarly distributed in multiple ranges, and high error was recorded (e.g., MAE: 6).
In contrast, road 16 and road 46 (Fig.~\ref{fig_teaser} B4) had low standard deviation value (e.g., 3) with a pointy and unbalanced speed distribution and a low error (e.g., MAE: 2).

Then, she began further investigation on the reason for such a relation, assuming that \textbf{(Finding 1) the model makes inaccurate predictions (i.e., high error) for roads with high speed fluctuations (i.e., large speed changes)}.
During her investigation, she found many cases that could confirm this assumption.
For example, Fig.~\ref{fig_teaser}C shows two example roads--road 81: high fluctuation and, road 18: low fluctuation.
At road 81, the model did not accurately forecast when the speed suddenly dropped or soared (MAE: 6), but it accurately predicted speeds for road 18 (MAE: 2).
She thought this finding could also explain why there are many roads with higher error in the intersections of regions (IR) that experience severe speed changes \textbf{(Q1)}.

As she understood which roads tend to have high error and when the high error occurs, she investigated how the model makes prediction with the roads and possible causes of inaccuracy\textbf{(Q2--Q4)}. 
To do so, she first set the attention filter (Fig.~\ref{fig_teaser} A2) as 0.1 to pass the roads that the model did not give much attention.
Then, she clicked road 113, one of the highlighted roads with the highest error on the map (Fig.~\ref{fig_ulsancase}B, yellow-fill circle) to add it to the line view and turned on \ar. 
Once the speed data line was added, she moved the blue bar in the line view to the point that had a low absolute error (Fig.~\ref{fig_ulsan_sptn} T1), updating all other visualization views. 
Based on the updated map and attention views, \textbf{(Q4)} she found that the model mainly refers to road 113 and road 86 \rev{(}Fig.~\ref{fig_ulsancase} A).
She also noticed several vertical light--pink lines in the attention view, including that of road 113 and 86. 
The reference to road 113 is a self-reference, implying that it is the model used in the road's past speed data (12 steps, 1 hour) for forecasting \textbf{(Q3)}. 
This self-reference at the low speed deviation is advantageous when the current road speed follows a past speed trend, or periodic pattern, because the speed pattern does not change much from the previous steps~\cite{yao2019revisiting}.

It is interesting that the model referred to road 86 among many other candidates.  
Initially, she confirmed that road 86's speed trend was a similar to that of road 113 (Fig.~\ref{fig_ulsan_sptn} A). 
In addition, she found that the model effectively attended road 86's information.  
For example, there was a time when road 86 had a speed drop (Fig.~\ref{fig_ulsan_sptn} P1), which is different from its average speed trend. 
When this sudden drop occurred, the model did not refer to road 86 (Fig.~\ref{fig_ulsancase} P1) as it learned the temporal dependency that referring to road 86 in the drop timing is not helpful for accurate predictions for road 113.

Once she determined how the model makes accurate predictions, she became interested in why the model behaves differently for the road with a high speed fluctuation, making an inaccurate prediction \textbf{(Q2--Q4)}. 
As she moved the blue bar in the line view to the point where road 113 had a sudden speed change and high error (Fig.~\ref{fig_ulsan_sptn}, T2), she found that the model began attending other roads (e.g., road 39, 54, 86) and reduced existing self-reference intensity (Fig.~\ref{fig_ulsancase} B, attention view). 
The head-cluster view also showed similar behaviors of the model in terms of cluster point of view. 
For example, when the model has sufficient self-reference, there are dark red diagonal patterns, similar to Head 2, 3, and 4 in the low error case (Fig.~\ref{fig_headcluster} top right).
However, when the model lacks self-reference, these patterns vanish from the matrix, as shown in the high error case (Fig.~\ref{fig_headcluster}, top left).

She speculated that this attention behavior is not helpful for an accurate prediction of road 113 for two reasons. 
First, as the referred roads (e.g., road 39, 54 and 86) are far from the target road (Fig.~\ref{fig_ulsancase}B), there is a low probability that the referred roads have similar traffic patterns as that of road 113 and any congestion is propagated among the roads in the near-future (e.g., 15 minutes) with given speed levels~\cite{guo2019identifying}. 
Second, a distant road can be helpful for prediction if there is any relationship between the traffic patterns of the roads.
For example, if the traffic pattern of a referred road precedes that of a target road, referring to distant roads could be advantageous~\cite{diks2006new}. 
But, she could not find crucial evidence of preceding traffic patterns from the line chart view.
She also noticed from the cluster visualization and table view they are not even in the same cluster, as shown Fig.~\ref{fig_teaser}B.

Rather, during this investigation, she found that road 112 is close to road 113 and is also in the same cluster as road 113. 
The speed pattern also seeded to precede that of Road 113, as shown in Fig.~\ref{fig_ulsan_sptn} P2.
To inspect this further, she ran a Granger causality test~\cite{diks2006new} with Road 112 and 113. 
The test result indicated that road 112's speed trend indeed precedes that of road 113 (F[6,268]=16.2, p=0.001), as shown Fig.~\ref{fig_ulsancase} B1 (Table). 
These observations led her to think that \textbf{(Finding 2) the model effectively captures similar speed trends, but may not effectively recognize preceding speed patterns (Q2).}
She also noted that \textbf{(Finding 3) in the case of high speed fluctuations, the model may attend many other roads and lose the importance of self-reference (Q2, Q4).}
This behavior is also not appropriate, as existing autoregressive models have already shown that past temporal data acquired by self-reference is more crucial for short-term forecasting (e.g., 15 minutes)~\cite{chen2011short, madzlan2010arima}.

\begin{figure}[t]
  \centering
  \includegraphics[width=0.85\columnwidth]{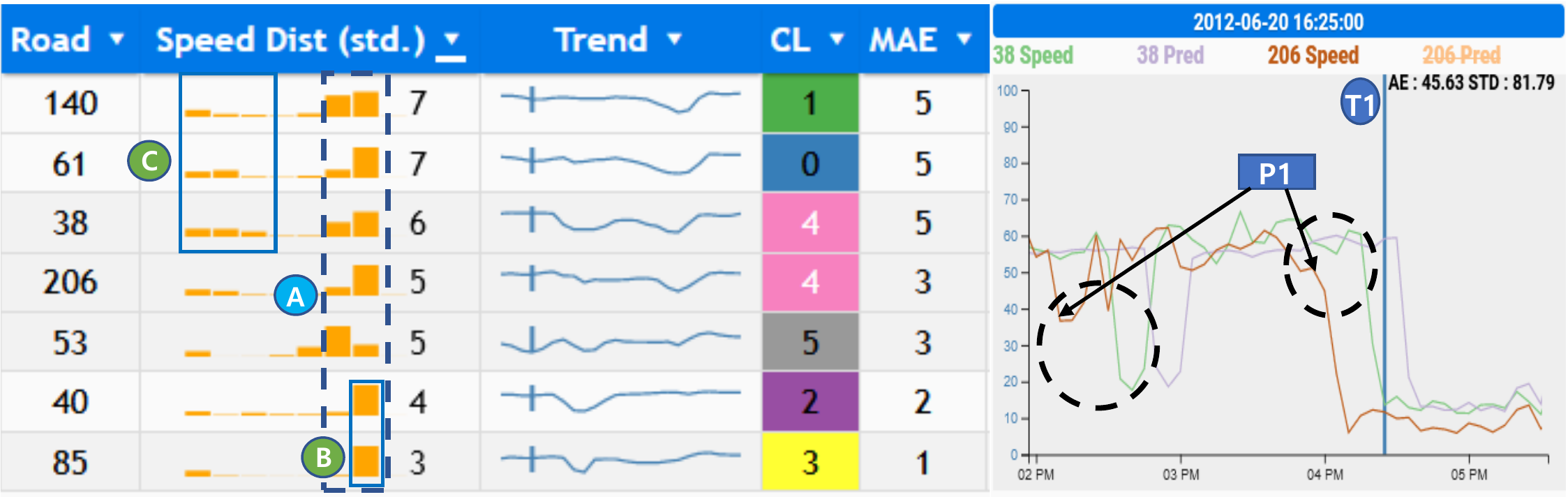}
   \vspace{-0.3cm}
  \caption{Summary of tables and Causality Analysis in LA. Overall there is a positive correlation between std. and MAE. Road 206's traffic pattern precedes that of road 38 (line chart, P1). }
  \label{fig_latable}
  \vspace{-0.7cm}
\end{figure}

\subsection{Analysis with Highway Roads}

To determine whether the model shows the same behaviors with different road networks \textbf{(Q6)}, she decided to analyze the model with the speed data on highways in L.A., USA.
As her purpose includes confirming the model behaviors on the roads with high and low errors \textbf{(Q1, Q2)}, the analysis procedure is similar to that of Ulsan in the previous study. After she replaced existing data with L.A. data in the menu (Fig.~\ref{fig_teaser}A), she set the MAE filter as 3.6--the value at the third quartile and sorted the table view by the speed distribution.  
Fig.~\ref{fig_latable} (left) shows an example sorted table by the distribution.

She observed that most roads have similar speed trends across the clusters. 
This could be the characteristic of highway roads since the speed is rather faster than the urban road networks.
Speed distributions are skewed right and pointy in the table (Fig.~\ref{fig_latable}A), indicating typical highways' speed trends.
She also obtained an overview of the positive correlation between standard deviation and MAE from the table.
For example, roads 140, 61, 38 had a high error having wide range of speed (Fig.~\ref{fig_latable}C), while roads 40 and 85 with a pointy speed distribution had a low error (Fig.~\ref{fig_latable}B), which confirms \textbf{(Finding 1)}. 
She also noticed from the map view that high error roads exist at the intersections of regions (IR) and edges of regions (ER)~\cite{la_city} (Fig.~\ref{fig_lacase}A, roads with black dots inside).
This pattern was also found in the previous study.

\begin{figure}
  \centering
  \includegraphics[width=1\columnwidth]{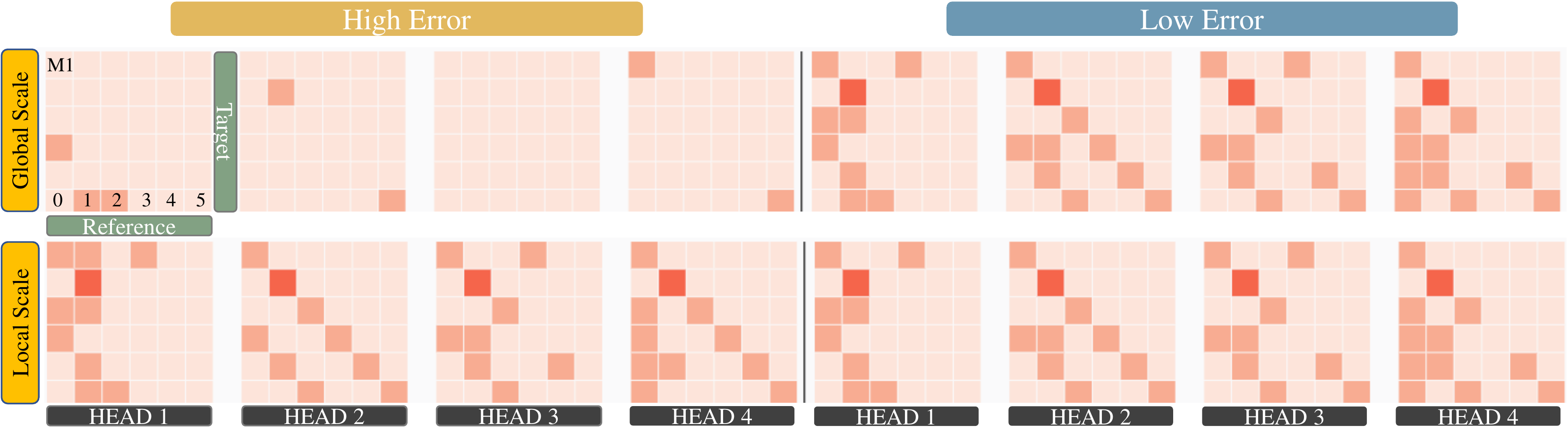}
  \caption{ The head-cluster view with four attention heads (top) in LA.
  }
  \label{fig_laheadcluster}
\end{figure}

As she clicked on road 38 on the map view, one of the black dots (i.e., high error), she moved the blue bar to the point where the model recorded a high error (e.g., AE: 45.63, Fig.~\ref{fig_latable} T1).
Then, she instantly noticed several light--pink vertical bars in the attention view (Fig.~\ref{fig_lacase}B, attention view) and found that none of the bars had an attention intensity larger than 0.1.
She thought that this was the same behavior shown in the previous study and may not be a proper behavior because the roads that encode preceding traffic patterns of other roads~\textbf{(Finding 2)} or refer its own past speed pattern~\textbf{(Finding 3)} tend to have high accuracy. 
The head-cluster view also showed similar behaviors from a cluster point of view (Appendix Fig.~\ref{fig_laheadcluster}). 
The head-cluster view displayed many light pink clusters in the diagonal pattern (i.e., insufficient self-reference), when the model recorded high error. 
However, with sufficient self-reference, it had high accuracy with dark red diagonal patterns.

To further inspect the model's attention behavior, she performed a causality analysis (CA) and found that road 206 has a speed pattern that precedes that of road 38 (i.e., dropping and increasing traffic speed earlier than road 38's traffic speed), as shown in Fig.~\ref{fig_latable} P1.
The Granger causality test confirmed that road 206's speed pattern indeed exceeds that of road 38 (F[5, 271]=5.9, p$<$0.001) as shown in Fig.~\ref{fig_lacase} B1 (Table).
But the model rarely attended road 206, nor performed self-referencing with attention intensity less than 0.1~\textbf{(Finding 2)}.
The pie chart of road 38 (yellow) and attention matrix in Fig.~\ref{fig_lacase}B presented this observations. 
Note that the model had performed self-referencing for road 38's prediction, before the traffic speed fluctuated (\textbf{Finding 3)}, as shown in Fig.~\ref{fig_lacase} A, pixel view.

\subsection{Attention Enforcement Tests}
\label{sec_enforce}

From the two case studies with \toolname, she had three main findings. \textbf{Finding 1}: when a road's speed highly fluctuates, the model's error increases; \textbf{Finding 2}: the model often fails to find appropriate references with preceding speed trends; and \textbf{Finding 3}: the model loses important past self-reference information by dispersing attention to other unimportant roads.

\begin{figure}[t]
  \centering
  \includegraphics[width=.95\columnwidth]{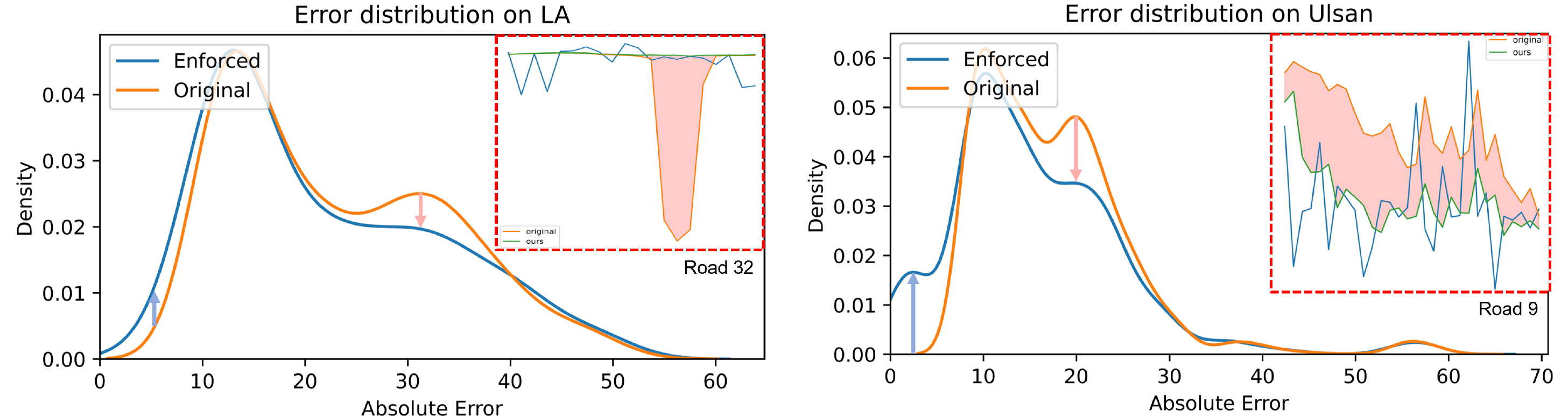}
  \vspace{-0.4cm}
  \caption{A result of the attention enforcement presents error distribution.}
  \label{fig_report}
  \vspace{-0.7cm}
\end{figure}

As a next step, she decided to improve the model performance with the findings, using the attention enforcement method, provided in the attention view (Sec.~\ref{sec_attn}). 
First, she selected four clusters in the diagonal pattern at Head 4, as shown in Fig.~\ref{fig_headcluster}, green dotted box. 
As she clicked the ``Test Alternatives'' button in the view, a new view popped up that presents two line charts for performance comparison (Fig.~\ref{fig_report}). 
From the chart, she observed a shift in the original model's chart, which indicates improved performance. 
For example, from the chart of the METR-LA data, she recognized that the number of roads with about 30 absolute error (AE) decreased, while those with about 5 AE increased. 
Similarly, the number of roads with about 20 AE decreased the same as those with about 5 AE increased with the Ulsan data. 
She thought this result came from the fact that the attention enforcement method had the model focus on self-reference and roads with preceding speed patterns in the same clusters, confirming and using \textbf{Finding 2 and 3} together.

\begin{table}[t]
\centering
\caption{Improved accuracy using the findings from our visual analytics approach. 10\% of roads with the highest error are used.
}
\resizebox{0.9\columnwidth}{!}{\begin{tabular}{c|c|cccc}
                                                Dataset & Method & 15 Mins & 30 Mins & 60 Mins & Average \\
                            \hline
\multirow{3}{*}{{Ulsan}} & \stgrat & 8.70     & 9.00     & 9.52     & 8.98    \\
                             &  Enforced w/ DTW & \textbf{8.23}     & 8.43     & 8.81     & 8.39  \\
                             &  Enforced w/ DTW + Granger & \textbf{8.24}     & \textbf{8.37}     & \textbf{8.76}     & \textbf{8.37}  \\
\hline
\multirow{3}{*}{{METR-LA}}  & \stgrat & 6.08 & 7.43 & 10.61 & 7.68    \\
                             & Enforced w/ DTW & 6.06     & 7.38     & 10.24    & 7.55  \\
                             & Enforced w/ DTW + Granger & \textbf{6.04}     & \textbf{7.28}     & \textbf{10.08}    & \textbf{7.48}  \\
\hline
\end{tabular}}
\label{table.AttnEnforcement}
\vspace{-0.7cm}
\end{table}

\section{Expert Feedback}
\label{sec_feedback}
To evaluate our VA approach, we conducted a semi-structured interview session \rev{(2-hour long)} with two domain experts, E3 and E4. 
E3 and E4 had 6 and 2 years of experience, respectively, as machine learning engineers in the traffic domain, developing deep learning models in the transportation domain.  
E3 had also participated in meetings to perform task analysis (Sec.~\ref{task_analysis}). 
Before the session, we performed an online tutorial session on \toolname and sent them online access to the system so that they could freely use and test the system. 
In the interview session, we asked about their challenges at work, showcased the system with the case studies, and had a Q\&A session on the system. 
We then asked about the usefulness and impact of the system on their work.

First, when we asked them about the challenges in their work, they explained that improving models in the traffic domain is difficult, because the data have complicated spatio-temporal dependencies, and they do not have effective tools for analyzing the dependencies in the layers. 
Thus, the process is performed in a brute-force manner using scripting languages-- 
\textit{``...whenever we evaluate the importance of some features or layers, we have to iteratively retrain our model and check the raw outputs,''} E4 reported.
E4 also commented that the attention view and AttnArrows are helpful for analyzing complex spatio-temporal attention, and they can easily discover important features and layers and debug models' misendeavor in advance, which is critical for shortening their work time due to the reduced number of retraining in the end.

E3 expressed similar views to E4 regarding the system's usefulness for analyzing attention, as well as exploring and debugging model behaviors.
In particular, he mentioned that the system is effective in cases, where DL models do not have specific performance improvement, such as speed predictions for rush hours~\cite{Lee19, Lee21}--\textit{``...there are cases where deep learning models' performance do not outperform existing approaches. ...using the attention enforcement, we can have improved performance, better responding to user requests with higher accuracy.''}

He then commented that the system allows great inspiration and has large potential in optimizing computing resources in a novel way.
He explained that as deep learning models become heavier with many layers and input features, consuming many computing resources, it becomes a burden to utilize heavy models all the time.
In this case, he suggested using the system at work with two types of models--one with high accuracy and resource-demanding and another with reasonable accuracy without resource-demanding. 
Then, the attention mechanism is used to decide which model to use for prediction with consideration of resource usage (i.e., switching models for prediction by conditions). 
Then they can explore the behaviors of the attention as described in the case studies and even improve performance by using the attention enforcement method, all of which could contribute to saving computing resources--\textit{``...if we can interchange between lightweight and heavy models, [...] , we can save tremendous resources for our navigation and prediction services.''}

E3 gave positive feedback on the automated methods. 
According to E3, DTW and Granger causality methods have already been used and have allowed notable results in their teams dealing with time series data analysis and prediction. 
E3 said, \textit{``...DTW and the enforcement of attention with DTW is a very interesting and novel approach for the traffic forecasting.''}
Lastly E3 spotted that the system can be easily extended for other cities, which is advantageous, as they deal with many large cities for services--\textit{``Confirming models with different cities is also important [...] and this system can be frequently used for confirming model performance with many cities' data. ''}

Lastly, when asked how to strengthen the system, they answered that temporal analysis on the map could be further strengthened, such as direct time-series pattern visualization on the map with respect to roads, which could help developers better explore the dependencies. They also expressed their interest in using the system with DL models for travel time prediction, as it could reveal new types of insights into how deep learning models deal internally with sub-paths and local roads.

\section{Limitations, Discussion, and Future Work}
We encode clusters using colors because our algorithm experiment results show that there are less than eight meaningful traffic patterns in our urban and highway traffic data due to the inherent periodicity of the traffic on roads (e.g., existence of rush hour). 
Note that Kwon et al.~\cite{kwon18clustervision} have demonstrated that a greater number of clusters can be identified with colors (e.g., 20 clusters). 
Although we assign colors to best distinguish the clusters with reduced visual complexity, assigning colors to clusters based on similarity levels could allow for effective cluster analysis. 
For example, clusters with high similarities could have a similar color~\cite{kumar2019clustered}.
By modifying Eq.~\ref{eq_attn}, the attention matrix can be calculated from other attention-based spatio-temporal forecasting models. 
As Table~\ref{table.AttnEnforcement} with top 10\% high error roads shows, the attention enforcement method results in improved accuracy.  
While a negative relationship is known to exist between traffic volume and speed~\cite{andrienko2016leveraging, lana2018road, li2018brief}, it has not been investigated in this work; the model behavior with regard to this relationship could be the topic of a future study. The volume data for the LA dataset can be downloaded from the official website~\cite{division_2021}.
We place road information in the table view, but some information could be shown with visualization on the map for effective visual analysis, though this would also increase visual complexity.

We perform two experiments and acquire the best performance, when both DTW and Granger causality test are applied.
Note that the improvement is meaningful, as most recent traffic forecasting models are recognized with a similar level of performance improvements~\cite{Lee21}. 
The colored parts of the two inset visualizations in Fig.~\ref{fig_report} mean improved performance gaps and some roads show significant performance improvement (e.g., road 32, 9).
We show the usefulness of DTW and Granger causality tests for spatio-temporal attention analysis, but incorporating other methods (e.g, Bayesian inferences) also could produce additional insights.
\rev{We use a fixed range of error using quartiles for the sake of simplicity; hence, the high (Q3) and low (Q1) error bounds are relative and different datasets would have different low and high error bounds with different error distributions. 
A filtering function for setting the error bound threshold could help users effectively investigate model behaviors with different error ranges.}
\rev{Although the experts in the interview have not reported any concerns regarding the usability of the system, a user study could be performed to investigate usability and complexity issues.}
\rev{For in-depth analysis, a longitudinal study~\cite{saraiya2006insight} can be performed with the experts to show how the system can be used at work and to measure the impact of the system. }

As a future work, we plan to research how to make an automated switch based on attention between the light weight and heavy forecasting models for DL resource optimization. 
It is of interest regarding what VA systems can support the exploration of such models to answer when and how the models are switched during the forecasting process.

\section{Conclusion}
We design a VA approach to help users explore the process of traffic forecasting and improve model performance. 
We perform task analysis with domain experts, which inform our system design.
The system provides users with multiple views, including filter, line, map, and attention views, for effective model exploration in a spatio-temporal perspective. 
For evaluation, we perform two case studies, showing how users form and validate hypotheses and generate insights into model behaviors.
We also show that the insights derived by using our VA approach are critical in improving the accuracy.

%\acknowledgments{
%Fill the contents or delete this section
%}

\acknowledgments{
This work was supported by the Korean National Research Foundation (NRF) grant (No. 2021R1A2C1004542) and by the Institute of Information \& Communications Technology Planning \& Evaluation (IITP) grants (No. 2020-0-01336--Artificial Intelligence Graduate School Program, UNIST), funded by the Korea government (MSIT). This work was also partly supported by NAVER Corporation.}

\bibliographystyle{abbrv-doi}
\bibliography{ms}

\begin{thebibliography}{10}

\bibitem{afzalan2019automated}
M.~Afzalan and F.~Jazizadeh.
\newblock An automated spectral clustering for multi-scale data.
\newblock {\em Neurocomputing}, 347:94--108, 2019.

\bibitem{akhtar2021review}
M.~Akhtar and S.~Moridpour.
\newblock A review of traffic congestion prediction using artificial
  intelligence.
\newblock {\em Journal of Advanced Transportation}, 2021:1--18, 2021.

\bibitem{andrienko2003exploratory}
N.~Andrienko, G.~Andrienko, and P.~Gatalsky.
\newblock Exploratory spatio-temporal visualization: an analytical review.
\newblock {\em Journal of Visual Languages \& Computing}, 14(6):503--541, 2003.

\bibitem{andrienko2016leveraging}
N.~Andrienko, G.~Andrienko, and S.~Rinzivillo.
\newblock Leveraging spatial abstraction in traffic analysis and forecasting
  with visual analytics.
\newblock {\em Information Systems}, 57:172--194, 2016.

\bibitem{aven2017daily}
M.~Aven.
\newblock Daily traffic flow pattern recognition by spectral clustering.
\newblock {\em CMC Senior Theses}, p. 1597, 2017.

\bibitem{berndt1994using}
D.~J. Berndt and J.~Clifford.
\newblock Using dynamic time warping to find patterns in time series.
\newblock In {\em Proceedings of the International Conference on Knowledge
  Discovery and Data Mining}, p. 359–370, 1994.

\bibitem{cabrera2019fairvis}
A.~A. Cabrera, W.~Epperson, F.~Hohman, M.~Kahng, J.~Morgenstern, and D.~H.
  Chau.
\newblock Fairvis: Visual analytics for discovering intersectional bias in
  machine learning.
\newblock In {\em IEEE Conference on Visual Analytics Science and Technology},
  pp. 46--56, 2019.

\bibitem{chen2011short}
C.~Chen, J.~Hu, Q.~Meng, and Y.~Zhang.
\newblock Short-time traffic flow prediction with arima-garch model.
\newblock In {\em 2011 IEEE Intelligent Vehicles Symposium (IV)}, pp. 607--612,
  2011.

\bibitem{chen2012retrieval}
C.~Chen, Y.~Wang, L.~Li, J.~Hu, and Z.~Zhang.
\newblock The retrieval of intra-day trend and its influence on traffic
  prediction.
\newblock {\em Transportation research part C: emerging technologies},
  22:103--118, 2012.

\bibitem{revacnnfilm}
S.~Chung, C.~Park, S.~Suh, K.~Kang, J.~Choo, and B.~C. Kwon.
\newblock Revacnn: Steering convolutional neural network via real-time visual
  analytics.
\newblock {\em Future of Interactive Learning Machines Workshop(FILM at
  NerIPS)}, 2016.

\bibitem{collins2009bubble}
C.~Collins, G.~Penn, and S.~Carpendale.
\newblock Bubble sets: Revealing set relations with isocontours over existing
  visualizations.
\newblock {\em IEEE Transactions on Visualization and Computer Graphics},
  15(6):1009--1016, 2009.

\bibitem{derose2020attention}
J.~F. DeRose, J.~Wang, and M.~Berger.
\newblock Attention flows: Analyzing and comparing attention mechanisms in
  language models.
\newblock {\em IEEE Transactions on Visualization and Computer Graphics},
  27(2):1160--1170, 2021.

\bibitem{derrow2021eta}
A.~Derrow-Pinion, J.~She, D.~Wong, O.~Lange, T.~Hester, L.~Perez, M.~Nunkesser,
  S.~Lee, X.~Guo, B.~Wiltshire, et~al.
\newblock Eta prediction with graph neural networks in google maps.
\newblock In {\em Proceedings of the 30th ACM International Conference on
  Information \& Knowledge Management}, pp. 3767--3776, 2021.

\bibitem{diks2006new}
C.~Diks and V.~Panchenko.
\newblock A new statistic and practical guidelines for nonparametric granger
  causality testing.
\newblock {\em Journal of Economic Dynamics and Control}, 30(9-10):1647--1669,
  2006.

\bibitem{division_2021}
D.~C. Division.
\newblock Traffic count data: Lac open data,
  https://data.lacounty.gov/transportation/traffic-count-data/uvew-g569, Oct
  2021.

\bibitem{fang2020constgat}
X.~Fang, J.~Huang, F.~Wang, L.~Zeng, H.~Liang, and H.~Wang.
\newblock Constgat: Contextual spatial-temporal graph attention network for
  travel time estimation at baidu maps.
\newblock In {\em Proceedings of the 26th ACM SIGKDD International Conference
  on Knowledge Discovery \& Data Mining}, pp. 2697--2705, 2020.

\bibitem{whatif_google}
Google.
\newblock Model understanding with the what-if tool dashboard.
\newblock Available at
  \url{https://www.tensorflow.org/tensorboard/what_if_tool/}.

\bibitem{Guidotti18}
R.~Guidotti, A.~Monreale, S.~Ruggieri, F.~Turini, F.~Giannotti, and
  D.~Pedreschi.
\newblock A survey of methods for explaining black box models.
\newblock {\em ACM Comput. Surv.}, 51(5):1--42, 2018.

\bibitem{guo2019identifying}
S.~Guo, D.~Zhou, J.~Fan, Q.~Tong, T.~Zhu, W.~Lv, D.~Li, and S.~Havlin.
\newblock Identifying the most influential roads based on traffic correlation
  networks.
\newblock {\em EPJ Data Science}, 8(1):1--17, 2019.

\bibitem{harrower2003colorbrewer}
M.~Harrower and C.~A. Brewer.
\newblock Colorbrewer. org: an online tool for selecting colour schemes for
  maps.
\newblock {\em The Cartographic Journal}, 40(1):27--37, 2003.

\bibitem{Hochreiter97}
S.~Hochreiter and J.~Schmidhuber.
\newblock Long short-term memory.
\newblock {\em Neural Comput.}, 9(8):1735–1780, 1997.

\bibitem{hohman2019summit}
F.~{Hohman}, H.~{Park}, C.~{Robinson}, and D.~H. {Polo Chau}.
\newblock Summit: Scaling deep learning interpretability by visualizing
  activation and attribution summarizations.
\newblock {\em IEEE Transactions on Visualization and Computer Graphics},
  26(1):1096--1106, 2020.

\bibitem{hohman2018visual}
F.~M. Hohman, M.~Kahng, R.~Pienta, and D.~H. Chau.
\newblock Visual analytics in deep learning: An interrogative survey for the
  next frontiers.
\newblock {\em IEEE Transactions on Visualization and Computer Graphics},
  25(8):2674--2693, 2019.

\bibitem{Jagadish14}
H.~V. Jagadish, J.~Gehrke, A.~Labrinidis, Y.~Papakonstantinou, J.~M. Patel,
  R.~Ramakrishnan, and C.~Shahabi.
\newblock Big data and its technical challenges.
\newblock {\em Commun. ACM}, 57:86--94, 2014.

\bibitem{kahng18GanLab}
M.~{Kahng}, N.~{Thorat}, D.~H. {Chau}, F.~{Vi{\'e}gas}, and M.~{Wattenberg}.
\newblock {GAN Lab: Understanding Complex Deep Generative Models using
  Interactive Visual Experimentation}.
\newblock {\em IEEE Transactions on Visualization and Computer Graphics},
  25:310--320, 2019.

\bibitem{Kahng2017activis}
M.~Kahng, P.~Y.~Andrews, A.~Kalro, and D.~Horng Polo~Chau.
\newblock Activis: Visual exploration of industry-scale deep neural network
  models.
\newblock {\em IEEE Transactions on Visualization and Computer Graphics},
  24(1):88--97, 2017.

\bibitem{keim2000designing}
D.~A. Keim.
\newblock Designing pixel-oriented visualization techniques: Theory and
  applications.
\newblock {\em IEEE Transactions on Visualization and Computer Graphics},
  6(1):59--78, 2000.

\bibitem{Kenney11}
J.~B. {Kenney}.
\newblock Dedicated short-range communications (dsrc) standards in the united
  states.
\newblock {\em Proceedings of the IEEE}, 99(7):1162--1182, 2011.

\bibitem{kipf2016semi}
T.~N. Kipf and M.~Welling.
\newblock {Semi-Supervised Classification with Graph Convolutional Networks}.
\newblock In {\em Proceedings of the International Conference on Learning
  Representations}, 2017.

\bibitem{ko2014analyzing}
S.~Ko, S.~Afzal, S.~Walton, Y.~Yang, J.~Chae, A.~Malik, Y.~Jang, M.~Chen, and
  D.~Ebert.
\newblock Analyzing high-dimensional multivariate network links with integrated
  anomaly detection, highlighting and exploration.
\newblock In {\em IEEE Conference on Visual Analytics Science and Technology},
  pp. 83--92, 2014.

\bibitem{ko2012marketanalyzer}
S.~Ko, R.~Maciejewski, Y.~Jang, and D.~S. Ebert.
\newblock Marketanalyzer: An interactive visual analytics system for analyzing
  competitive advantage using point of sale data.
\newblock {\em Computer Graphics Forum}, 31(3pt3):1245--1254, 2012.

\bibitem{kumar2019clustered}
A.~Kumar, N.~Timmermans, M.~Burch, and K.~Mueller.
\newblock Clustered eye movement similarity matrices.
\newblock In {\em Proceedings of the 11th ACM Symposium on Eye Tracking
  Research \& Applications}, pp. 1--9, 2019.

\bibitem{kwon18clustervision}
B.~C. Kwon, B.~Eysenbach, J.~Verma, K.~Ng, C.~deFilippi, W.~F. Stewart, and
  A.~Perer.
\newblock Clustervision: Visual supervision of unsupervised clustering.
\newblock {\em IEEE Transactions on Visualization and Computer Graphics},
  24(1):142--151, 2018.

\bibitem{lana2018road}
I.~Lana, J.~Del~Ser, M.~Velez, and E.~I. Vlahogianni.
\newblock Road traffic forecasting: Recent advances and new challenges.
\newblock {\em IEEE Intelligent Transportation Systems Magazine},
  10(2):93--109, 2018.

\bibitem{le2019shape}
V.~Le~Guen and N.~Thome.
\newblock Shape and time distortion loss for training deep time series
  forecasting models.
\newblock {\em Advances in neural information processing systems}, 32, 2019.

\bibitem{Lee19}
C.~{Lee}, Y.~{Kim}, S.~{Jin}, D.~{Kim}, R.~{Maciejewski}, D.~{Ebert}, and
  S.~{Ko}.
\newblock A visual analytics system for exploring, monitoring, and forecasting
  road traffic congestion.
\newblock {\em IEEE Transactions on Visualization and Computer Graphics},
  26(11):3133--3146, 2020.

\bibitem{Lee21}
H.~Lee, C.~Park, S.~Jin, H.~Chu, J.~Choo, and S.~Ko.
\newblock An empirical experiment on deep learning models for predicting
  traffic data.
\newblock In {\em 2021 IEEE 37th International Conference on Data Engineering
  (ICDE)}, pp. 1817--1822, 2021.

\bibitem{li2015trend}
L.~Li, X.~Su, Y.~Zhang, Y.~Lin, and Z.~Li.
\newblock Trend modeling for traffic time series analysis: An integrated study.
\newblock {\em IEEE Transactions on Intelligent Transportation Systems},
  16(6):3430--3439, 2015.

\bibitem{li2018brief}
Y.~Li and C.~Shahabi.
\newblock A brief overview of machine learning methods for short-term traffic
  forecasting and future directions.
\newblock {\em SIGSPATIAL Special}, 10(1):3--9, 2018.

\bibitem{li2018dcrnn}
Y.~Li, R.~Yu, C.~Shahabi, and Y.~Liu.
\newblock Diffusion convolutional recurrent neural network: Data-driven traffic
  forecasting.
\newblock In {\em Proceedings of the International Conference on Learning
  Representations}, pp. 1--16, 2018.

\bibitem{liu2018analyzing}
M.~Liu, S.~Liu, H.~Su, K.~Cao, and J.~Zhu.
\newblock Analyzing the noise robustness of deep neural networks.
\newblock In {\em IEEE Conference on Visual Analytics Science and Technology},
  pp. 60--71, 2018.

\bibitem{liu16cnnvis}
M.~Liu, J.~Shi, Y.~Li, C.~Li, J.~Zhu, and S.~Liu.
\newblock Towards better analysis of deep convolutional neural networks.
\newblock {\em IEEE Transactions on Visualization and Computer Graphics},
  23:91--100, 2016.

\bibitem{la_city}
{Los Angeles City Planning}.
\newblock Citywide maps.
\newblock Available at \url{https://planning.lacity.org/}.

\bibitem{madzlan2010arima}
N.~Madzlan, K.~Ibrahim, et~al.
\newblock Arima models for bus travel time prediction.
\newblock {\em Journal of the Institution of Engineers Malaysia}, 2010.

\bibitem{meulemans2013kelpfusion}
W.~Meulemans, N.~H. Riche, B.~Speckmann, B.~Alper, and T.~Dwyer.
\newblock Kelpfusion: A hybrid set visualization technique.
\newblock {\em IEEE Transactions on Visualization and Computer Graphics},
  19(11):1846--1858, 2013.

\bibitem{ming2017understanding}
Y.~Ming, S.~Cao, R.~Zhang, Z.~Li, Y.~Chen, Y.~Song, and H.~Qu.
\newblock Understanding hidden memories of recurrent neural networks.
\newblock In {\em IEEE Conference on Visual Analytics Science and Technology},
  pp. 13--24, 2017.

\bibitem{miyahara2014family}
S.~Miyahara and S.~Miyamoto.
\newblock A family of algorithms using spectral clustering and dbscan.
\newblock In {\em IEEE International Conference on Granular Computing}, pp.
  196--200, 2014.

\bibitem{Muhlbacher14}
T.~{Mühlbacher}, H.~{Piringer}, S.~{Gratzl}, M.~{Sedlmair}, and M.~{Streit}.
\newblock Opening the black box: Strategies for increased user involvement in
  existing algorithm implementations.
\newblock {\em IEEE Transactions on Visualization and Computer Graphics},
  20(12):1643--1652, 2014.

\bibitem{whatif_pair}
PAIR.
\newblock What-if tool.
\newblock Available at \url{https://pair-code.github.io/what-if-tool/}.

\bibitem{Pan2018HyperSTNetHF}
Z.~Pan, Y.~Liang, J.~Zhang, X.~Yi, Y.~Yu, and Y.~Zheng.
\newblock Hyperst-net: Hypernetworks for spatio-temporal forecasting.
\newblock {\em ArXiv}, abs/1809.10889, 2018.

\bibitem{park2019stgrat}
C.~Park, C.~Lee, H.~Bahng, Y.~Tae, S.~Jin, K.~Kim, S.~Ko, and J.~Choo.
\newblock St-grat: A novel spatio-temporal graph attention networks for
  accurately forecasting dynamically changing road speed.
\newblock In {\em Proceedings of the ACM International Conference on
  Information \& Knowledge Management}, p. 1215–1224, 2020.

\bibitem{cbpark2019sanvis}
C.~Park, I.~Na, Y.~Jo, S.~Shin, J.~Yoo, B.~C. Kwon, J.~Zhao, H.~Noh, Y.~Lee,
  and J.~Choo.
\newblock Sanvis: Visual analytics for understanding self-attention networks.
\newblock {\em IEEE Conference on Visual Analytics Science and Technology
  Short}, 2019.

\bibitem{pensky2019spectral}
M.~Pensky and T.~Zhang.
\newblock Spectral clustering in the dynamic stochastic block model.
\newblock {\em Electronic Journal of Statistics}, 13(1):678--709, 2019.

\bibitem{pezzotti2018deepeyes}
N.~Pezzotti, T.~Hollt, J.~Van~Gemert, B.~Lelieveldt, E.~Eisemann, and
  A.~Vilanova.
\newblock Deepeyes: Progressive visual analytics for designing deep neural
  networks.
\newblock {\em IEEE Transactions on Visualization and Computer Graphics},
  24(1):98, 2018.

\bibitem{Pi19}
M.~{Pi}, H.~{Yeon}, H.~{Son}, and Y.~{Jang}.
\newblock Visual cause analytics for traffic congestion.
\newblock {\em IEEE Transactions on Visualization and Computer Graphics},
  27(3):2186--2201, 2021.

\bibitem{Pu13}
J.~Pu, S.~Liu, Y.~Ding, H.~Qu, and L.~Ni.
\newblock T-watcher: A new visual analytic system for effective traffic
  surveillance.
\newblock In {\em IEEE International Conference on Mobile Data Management},
  vol.~1, pp. 127--136, 2013.

\bibitem{salvador2007toward}
S.~Salvador and P.~Chan.
\newblock Toward accurate dynamic time warping in linear time and space.
\newblock {\em Intelligent Data Analysis}, 11(5):561--580, 2007.

\bibitem{saraiya2006insight}
P.~Saraiya, C.~North, V.~Lam, and K.~A. Duca.
\newblock An insight-based longitudinal study of visual analytics.
\newblock {\em IEEE Transactions on Visualization and Computer Graphics},
  12(6):1511--1522, 2006.

\bibitem{shen2020visual}
Q.~Shen, Y.~Wu, Y.~Jiang, W.~Zeng, K.~Alexis, A.~Vianova, and H.~Qu.
\newblock Visual interpretation of recurrent neural network on
  multi-dimensional time-series forecast.
\newblock In {\em 2020 IEEE Pacific Visualization Symposium (PacificVis)}, pp.
  61--70. IEEE, 2020.

\bibitem{strobelt_seq2seq-vis_2019}
H.~Strobelt, S.~Gehrmann, M.~Behrisch, A.~Perer, H.~Pfister, and A.~M. Rush.
\newblock {SEQ}2seq-{VIS} : {A} {Visual} {Debugging} {Tool} for
  {Sequence}-to-{Sequence} {Models}.
\newblock {\em IEEE Transactions on Visualization and Computer Graphics},
  25(1):353--363, 2019.

\bibitem{strobelt2017lstmvis}
H.~Strobelt, S.~Gehrmann, H.~Pfister, and A.~M. Rush.
\newblock Lstmvis: A tool for visual analysis of hidden state dynamics in
  recurrent neural networks.
\newblock {\em IEEE Transactions on Visualization and Computer Graphics},
  24(1):667--676, 2017.

\bibitem{tedjopurnomo2020survey}
D.~A. Tedjopurnomo, Z.~Bao, B.~Zheng, F.~Choudhury, and A.~Qin.
\newblock A survey on modern deep neural network for traffic prediction:
  Trends, methods and challenges.
\newblock {\em IEEE Transactions on Knowledge and Data Engineering},
  34(4):1544--1561, 2022.

\bibitem{Thomas05}
J.~Thomas and K.~Cook.
\newblock {\em Illuminating the Path: The Research and Development Agenda for
  Visual Analytics}.
\newblock National Visualization and Analytics Ctr, 2005.

\bibitem{tufte2006beautiful}
E.~R. Tufte.
\newblock {\em Beautiful evidence}.
\newblock Graphis Pr, 2006.

\bibitem{vaswani2017attention}
A.~Vaswani, N.~Shazeer, N.~Parmar, J.~Uszkoreit, L.~Jones, A.~N. Gomez,
  {\L}.~Kaiser, and I.~Polosukhin.
\newblock Attention is all you need.
\newblock In {\em Advances in neural information processing systems}, vol.~30,
  pp. 5998--6008, 2017.

\bibitem{Velickovic2018graph}
P.~Veličković, G.~Cucurull, A.~Casanova, A.~Romero, P.~Liò, and Y.~Bengio.
\newblock Graph attention networks.
\newblock In {\em International Conference on Learning Representations}, 2018.

\bibitem{wang2019dqnvis}
J.~{Wang}, L.~{Gou}, H.~{Shen}, and H.~{Yang}.
\newblock Dqnviz: A visual analytics approach to understand deep q-networks.
\newblock {\em IEEE Transactions on Visualization and Computer Graphics},
  25(1):288--298, 2019.

\bibitem{wang2018GANViz}
J.~Wang, L.~Gou, H.~Yang, and H.-W. Shen.
\newblock Ganviz: A visual analytics approach to understand the adversarial
  game.
\newblock {\em IEEE Transactions on Visualization and Computer Graphics},
  24:1905--1917, 2018.

\bibitem{Wang13}
Z.~Wang, M.~Lu, X.~Yuan, J.~Zhang, and H.~Van De~Wetering.
\newblock Visual traffic jam analysis based on trajectory data.
\newblock {\em IEEE Transactions on Visualization and Computer Graphics},
  19(12):2159--2168, 2013.

\bibitem{Wexler20}
J.~{Wexler}, M.~{Pushkarna}, T.~{Bolukbasi}, M.~{Wattenberg}, F.~{Viégas}, and
  J.~{Wilson}.
\newblock The what-if tool: Interactive probing of machine learning models.
\newblock {\em IEEE Transactions on Visualization and Computer Graphics},
  26(1):56--65, 2020.

\bibitem{wu2020connecting}
Z.~Wu, S.~Pan, G.~Long, J.~Jiang, X.~Chang, and C.~Zhang.
\newblock Connecting the dots: Multivariate time series forecasting with graph
  neural networks.
\newblock In {\em Proceedings of the 26th ACM SIGKDD International Conference
  on Knowledge Discovery \& Data Mining}, pp. 753--763, 2020.

\bibitem{Wu19}
Z.~Wu, S.~Pan, G.~Long, J.~Jiang, and C.~Zhang.
\newblock Graph wavenet for deep spatial-temporal graph modeling.
\newblock In {\em Proceedings of the International Joint Conference on
  Artificial Intelligence}, pp. 1907--1913, 2019.

\bibitem{xiao2016traffic}
P.~Xiao, N.~Liu, Y.~Li, Y.~Lu, X.-j. Tang, H.-w. Wang, and M.-x. Li.
\newblock A traffic classification method with spectral clustering in sdn.
\newblock In {\em 2016 17th International Conference on Parallel and
  Distributed Computing, Applications and Technologies (PDCAT)}, pp. 391--394,
  2016.

\bibitem{yao2019revisiting}
H.~Yao, X.~Tang, H.~Wei, G.~Zheng, and Z.~Li.
\newblock Revisiting spatial-temporal similarity: A deep learning framework for
  traffic prediction.
\newblock In {\em Proceedings of the AAAI Conference on Artificial
  Intelligence}, vol.~33, pp. 5668--5675, 2019.

\bibitem{yin2021deep}
X.~Yin, G.~Wu, J.~Wei, Y.~Shen, H.~Qi, and B.~Yin.
\newblock Deep learning on traffic prediction: Methods, analysis and future
  directions.
\newblock {\em IEEE Transactions on Intelligent Transportation Systems}, pp.
  1--17, 2021.

\bibitem{yosinski2015deepviz}
J.~Yosinski, J.~Clune, A.~Nguyen, T.~Fuchs, and H.~Lipson.
\newblock Understanding neural networks through deep visualization.
\newblock In {\em Proceedings of the International Conference on Machine
  Learning}, 2015.

\bibitem{yu2018spatio}
B.~Yu, H.~Yin, and Z.~Zhu.
\newblock Spatio-temporal graph convolutional networks: A deep learning
  framework for traffic forecasting.
\newblock In {\em Proceedings of the International Joint Conference on
  Artificial Intelligence}, pp. 3634--3640, 2018.

\bibitem{yuan2018hetero}
Z.~Yuan, X.~Zhou, and T.~Yang.
\newblock Hetero-convlstm: A deep learning approach to traffic accident
  prediction on heterogeneous spatio-temporal data.
\newblock In {\em Proceedings of the 24th ACM SIGKDD International Conference
  on Knowledge Discovery \& Data Mining}, pp. 984--992, 2018.

\bibitem{zeiler2014visualizing}
M.~D. Zeiler and R.~Fergus.
\newblock Visualizing and understanding convolutional networks.
\newblock In {\em Proceedings of the European Conference on Computer Vision},
  pp. 818--833, 2014.

\bibitem{Zeng13}
W.~Zeng, C.-W. Fu, S.~M. Arisona, and H.~Qu.
\newblock Visualizing interchange patterns in massive movement data.
\newblock {\em Computer Graphics Forum}, 32(3--3):271--280, 2013.

\bibitem{zeng2020revisiting}
W.~Zeng, C.~Lin, J.~Lin, J.~Jiang, J.~Xia, C.~Turkay, and W.~Chen.
\newblock Revisiting the modifiable areal unit problem in deep traffic
  prediction with visual analytics.
\newblock {\em IEEE Transactions on Visualization and Computer Graphics},
  27(2):839--848, 2020.

\bibitem{Zhang2018gaan}
J.~Zhang, X.~Shi, J.~Xie, H.~Ma, I.~King, and D.~Yeung.
\newblock Gaan: Gated attention networks for learning on large and
  spatiotemporal graphs.
\newblock In {\em Proceedings of the Conference on Uncertainty in Artificial
  Intelligence}, pp. 339--349, 2018.

\end{thebibliography}

\onecolumn
\newpage
%\appendix
\section*{Appendix}

\begin{figure*}[ht]
    \centering \includegraphics[width=1\textwidth]{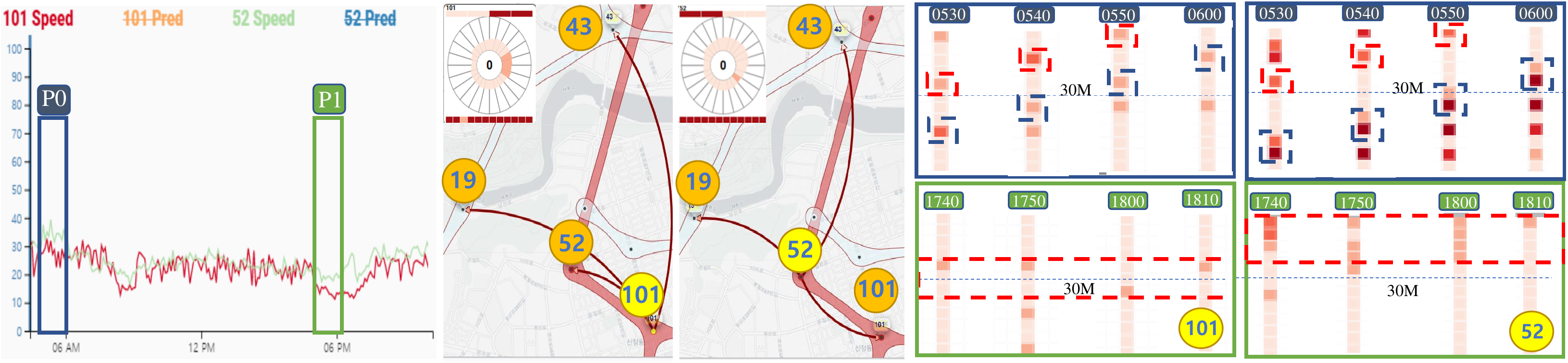}
    \caption{Pixel view shows temporal dependency tracking at different time periods.The two top enlarged matrix shows the pattern of P0, which tracks the sequences,  The two bottom shows the patterns of evening rush hour which monitors the fixed sequences.}
    \label{fig_pixelview}
\end{figure*}

\begin{table}[ht]\caption{Summary of experiment results on METR-LA dataset.}
\centering
\resizebox{\columnwidth}{!}{\begin{tabular}{l|l|lllll}
\hline
                    T    & Metric & GCRNN   & DCRNN   & STGCN   & Graph WaveNet   &  ST-GRAT\xspace                   \\ 
\hline
\multirow{3}{*}{15 min}  & MAE    & 2.80    & 2.73    & 2.88    & 2.69            & \textbf{2.60}      \\
                         & RMSE   & 5.51    & 5.27    & 5.74    & 5.15            & \textbf{5.07}      \\
                         & MAPE   & 7.5\%   & 6.99\%  & 7.62\%  & 6.90\%          & \textbf{6.61}\%    \\
\cline{1-7}
\multirow{3}{*}{30 min}  & MAE    & 3.24    & 3.12    & 3.47    & 3.07            & \textbf{3.01}      \\
                         & RMSE   & 6.74    & 6.36    & 7.24    & 6.26            & \textbf{6.21}      \\
                         & MAPE   & 9.0\%   & 8.65\%  & 9.57\%  & 8.37\%          & \textbf{8.15}\%   \\ 
\cline{1-7}
\multirow{3}{*}{1 hour}  & MAE    & 3.81    & 3.58    & 4.59    & 3.53            & \textbf{3.49}      \\
                         & RMSE   & 8.16    & 7.60    & 9.40    & \textbf{7.37}   & 7.42               \\
                         & MAPE   & 10.9\%  & 10.43\% & 12.70\% & \textbf{10.01}\%& \textbf{10.01}\%   \\ 
\cline{1-7}
\multirow{3}{*}{Average} & MAE    & 3.28    & 3.14    & 3.64    & 3.09            & \textbf{3.03}               \\
                         & RMSE   & 6.80    & 6.42    & 7.46    & 6.26            & \textbf{6.23}               \\
                         & MAPE   & 9.13\%  & 8.73\%  & 9.96\%  & 8.42\%          & \textbf{8.25}\%            \\
\cline{1-7}
\hline
\end{tabular}}
\label{table_result_metr}
\end{table}

\begin{figure}
  \centering
  \includegraphics[width=1\columnwidth]{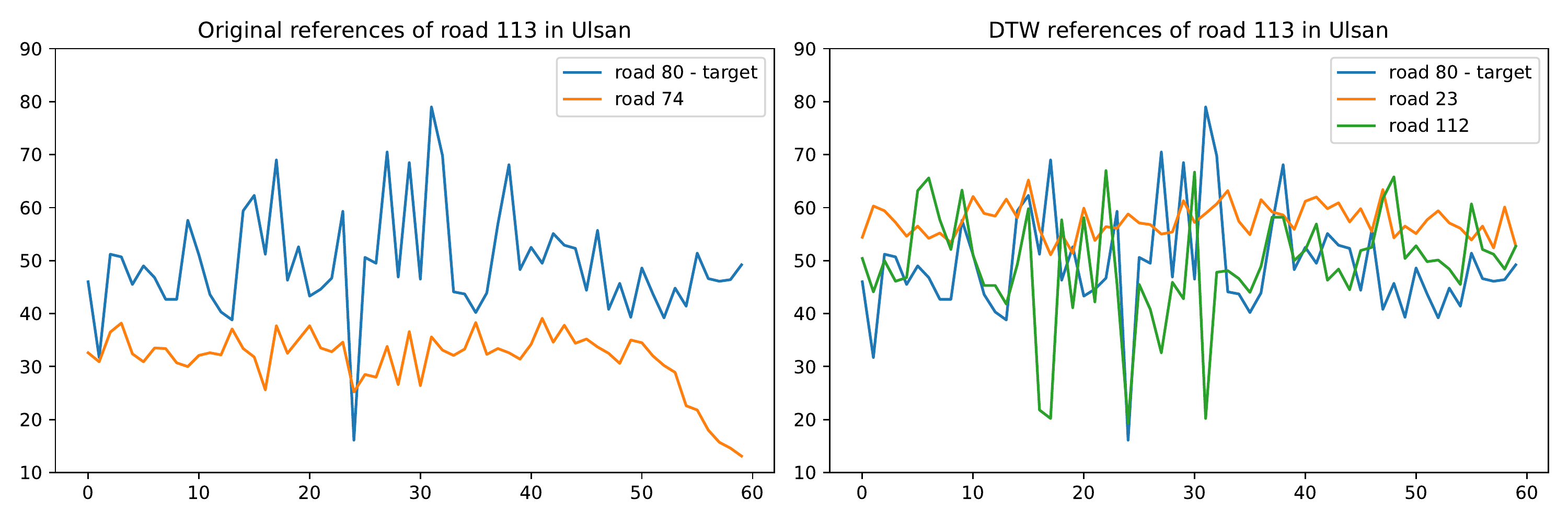}
  \caption{The comparison of the referring roads from spatial connection (i.e., neighboring roads) and from DTW in Ulsan}
  \label{fig_comp_ulsan_113}
\end{figure}

\begin{figure}
  \centering
  \includegraphics[width=1\columnwidth]{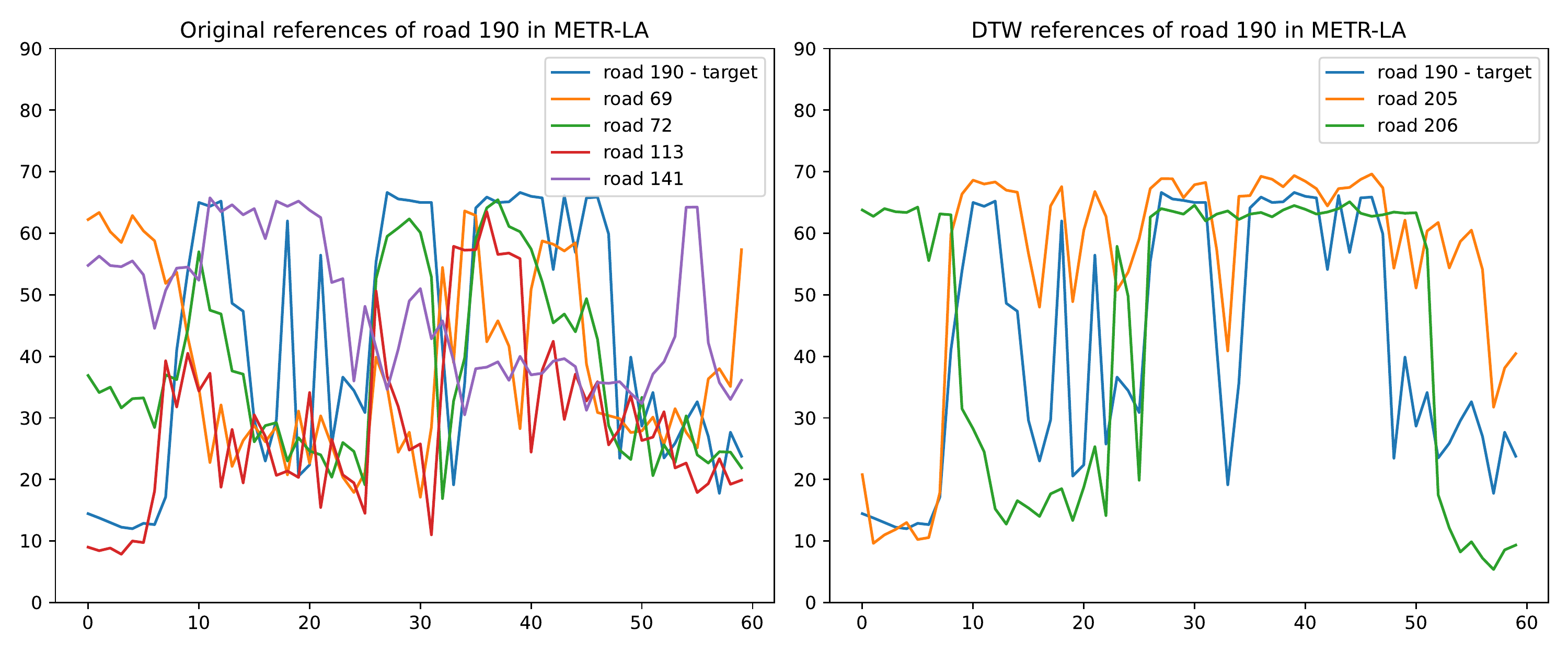}
  \caption{The comparison of the referring roads from spatial connection (i.e., neighboring roads) and from DTW in METR-LA}
  \label{fig_comp_metr_190}
\end{figure}
\end{document}